\documentclass[reprint,
twocolumn,
superscriptaddress,
nofootinbib,
 amsmath,amssymb,
 aps,
 5p,
 8pt
]{elsarticle}

\usepackage[utf8]{inputenc}
\usepackage{graphicx}% Include figure files
\usepackage{dcolumn}% Align table columns on decimal point
\usepackage{bm}% bold math
\usepackage[mathlines]{lineno}% Enable numbering of text and display math
\usepackage{csquotes}
\usepackage[T1]{fontenc}
\usepackage{amsmath} 
\usepackage{xcolor}
\usepackage{cuted}
\usepackage{mathabx}
\usepackage[normalem]{ulem}

\ifdefined\SAPO
\newcommand{\gda}[1]{}
\newcommand{\md}[1]{}
\newcommand{\mdc}[1]{}
\else
\newcommand{\gda}[1]{\textcolor{red}{#1}}

\newcommand{\md}[1]{\textcolor{pink}{#1}}
\newcommand{\mdc}[1]{\textcolor{pink}{[\textbf{MD:} #1]}}
\definecolor{azure}{rgb}{0.0, 0.5, 1.0}
\fi

\newcommand{\sv}{$\langle\sigma v \rangle$}

\usepackage[colorlinks=true, linkcolor=blue, citecolor=blue, urlcolor=blue]{hyperref}

\renewcommand\appendixautorefname[1]{}

\begin{document}

\title{Recasting and Forecasting Dark Matter Limits Without Raw Data: A Generalized Algorithm for Gamma-Ray Telescopes}

% Authors, for the paper (add full first names)
%\author{Giacomo D'Amico$^{1}$\orcidA{}*, Michele Doro$^{2,3}$\orcidB{}* and Michela de Caria$^{2}$}

% \author{Giacomo D'Amico}
% \affiliation{Institut de Fisica d'Altes Energies (IFAE), The Barcelona Institute of Science and Technology (BIST), E-08193 Bellaterra (Barcelona), Spain; gdamico@ifae.es}

% \author{Michele Doro}
% \affiliation{University of Padova, E-35131 Padova (PD), Italy; michele.doro@unipd.it }
% \affiliation{INFN sez. Padova, E-35131 Padova (PD), Italy; michele.doro@unipd.it}

% \author{Michela de Caria}
% \affiliation{University of Padova, E-35131 Padova (PD), Italy; }

\author[ifae]{Giacomo D'Amico\corref{cor1}}
\ead{gdamico@ifae.es}

\author[unipd,infn]{Michele Doro\corref{cor1}}
\ead{michele.doro@unipd.it}

\author[unipd]{Michela De Caria}

\cortext[cor1]{Corresponding author}

\address[ifae]{Institut de Física d'Altes Energies (IFAE), 
The Barcelona Institute of Science and Technology (BIST), 
E-08193 Bellaterra (Barcelona), Spain}

\address[unipd]{Department of Physics and Astronomy, University of Padova, I-35131 Padova (PD), Italy}

\address[infn]{INFN, Sezione di Padova, I-35131 Padova (PD), Italy}

%\longauthorlist{yes}

% MDPI internal command: Authors, for metadata in PDF
%\AuthorNames{Giacomo D'Amico, Michele Doro and Michela de Caria}

% MDPI internal command: Authors, for citation in the left column
%\AuthorCitation{D'Amico, G.; Doro, M.; de Caria, M.}
% If this is a Chicago style journal: Lastname, Firstname, Firstname Lastname, and Firstname Lastname.

% Affiliations / Addresses (Add [1] after \address if there is only one affiliation.)
%\address{%
%$^{1}$ \quad IFAE; giacomo.damico@ifae.es\\
%$^{2}$ \quad University of Padova, E-35131 Padova (PD), Italy; michele.doro@unipd.it \\
%$^{3}$ \quad INFN sez. Padova, E-35131 Padova (PD), Italy; michele.doro@unipd.it
%}

% Contact information of the corresponding author
%\corres{Correspondence: giacomo.damico@ifae.es (GDA); michele.doro@unipd.it (MD)}


\begin{abstract}

We present a novel  method for both forecasting and recasting upper limits (ULs) on dark matter (DM) annihilation cross sections, \(\left< \sigma v \right>^{UL}\), or decay lifetime $\tau^{LL}$ . The forecasting method relies solely on the instrument response functions (IRFs) to predict ULs for a given observational setup, without the need for full analysis pipelines.  The recasting  procedure uses published ULs to reinterpret constraints for alternative DM models or channels.  We demonstrate its utility across a range of canonical annihilation channels, including \(b\bar{b}\), \(W^+W^-\), \(\tau^+\tau^-\), and \(\mu^+\mu^-\), and apply it to several major gamma-ray experiments, including MAGIC, \textit{Fermi}-LAT, and CTAO. Notably, we develop a recasting approach that remains effective even when the IRF is unavailable by extracting generalized IRF-dependent coefficients from benchmark channels.  We apply this method to reinterpret ULs derived from standard spectra (e.g., PPPC4DMID) in terms of more recent DM scenarios, including a Higgsino-like model with mixed final states and spectra generated with the CosmiXs model. Extensive Monte Carlo simulations and direct comparison with published results confirm the robustness and accuracy of our method, with discrepancies remaining within statistical uncertainties. The algorithm is generally applicable to any scenario where the expected signal model is parametric, offering a powerful tool for reinterpreting existing gamma-ray limits and efficiently exploring the DM parameter space in current and future indirect detection experiments.

\end{abstract}

\maketitle

%\tableofcontents

\section{Introduction}
-
The nature of Dark Matter (DM) remains one of the most intriguing puzzles in modern Physics~\cite{Bergstrom:1997fj}. Despite its elusive nature, indirect search for DM, particularly through the detection of prompt high-energy cosmic messengers, like $\gamma$-rays, generated during DM annihilation or decay reactions, offer a promising route to constraining DM properties~\citep{Bergstrom:1997fj,Cirelli:2024ssz}. Indirect detection experiments searching for excess $\gamma$-ray signals from regions of high DM density, such as the Galactic center, the dwarf spheroidal galaxies, or galactic clusters, are conducted by satellite-based pair production instruments \citep{Atwood2009FermiLAT}, ground-based Imaging Atmospheric Cherenkov Telescopes (IACTs), such as HESS \citep{Hinton2004HESS}, MAGIC \citep{Aleksic2016MAGIC}, and VERITAS \citep{Holder2006VERITAS}, and in the future CTAO \citep{CTAConsortium2019Science}, as well as ground-based Shower Front Detectors (SFDs), such as HAWC \citep{Abeysekara2017HAWC}, LHAASO \citep{Cao2021LHAASO}, and in the future SWGO \citep{SWGO2022WhitePaper}, provide crucial insights into DM, especially in a  mass regime, above some TeVs, unattainable by direct detection experiments or colliders~\citep{Snowmass2013CosmicFrontierWorkingGroups1-4:2013wfj,GAMBIT:2017zdo}. While satellite-borne and SFDs are wide field-of-view (FOV) instruments, IACTs require targeted pointing. They have devoted thousands of hours in the previous decade in search for DM, hunting for several classes of targets~\cite{Doro:2021dzh,Hutten:2022hud}. 

Although the nature of DM is unknown, in the framework of Weakly Interacting Massive Particles (WIMPs), these can annihilate or decay. The velocity averaged annihilation cross-section, \(\left< \sigma v \right>\) or the decay lifetime $\tau$, along with the DM particle mass, are important experimental parameters in DM studies, as they relate to the interaction strength for annihilating DM particles of a given mass. 
%Setting upper limits (ULs) on \(\left< \sigma v \right>\) from observational data allows researchers to exclude certain DM models or at least constraint a specific DM annihilation scenarios. Similarly, limits can be extracted on the DM decay lifetime $\tau$. 
In either case, experimental limits are often computed for benchmark annihilation or decay channels, e.g. \(b\bar{b}\), \(\tau^+\tau^-\), \(W^+W^- \ldots\) translating the experimental counts limit into a \(\left< \sigma v \right>\) or $\tau$ limit. This requires knowledge of the instrument response functions (IRFs) such as the instrument effective area or energy reconstruction, as well as observational data such as the target observation time and sky trajectory. Reporting limits for a specific DM model, involves reconstructing the specific interaction mode across the allowed parameter space, and sacrifices generality. 
Until now, estimating limits for alternative benchmark channels or on specific DM models required a full experimental data reconstruction, usually lengthy for collaboration member and substantially impossible for external non-collaboration scientists who had a hard time assuming experiment performance~\citep[see, e.g.][]{Rodd:2024qsi}. 

In this work, we introduce a method that leverages the algebraic relationship between the expected number of signal events and the instrument's IRFs. This allows us to perform several predictions based on already published data: a) recast limits into alternative benchmark channels, b) recast them into specific DM models c) recast annihilation limits to decay limits d) forecast limits solely based on instrument IRFs. We tested the algorithm using both Monte Carlo simulations and published data. 
%
%By applying this framework, we are able to perform channel-to-channel predictions using existing datasets. This not only extends the utility of current experimental results but also allows researchers to impose constraints on theoretical models of DM that predict annihilation into multiple channels. Furthermore, t
The applicability is demonstrated on several of the mentioned instruments along with some further selected scenarios including the recast for different DM photon yields model, as the case for the recent results of \citet{arina2024cosmixs} or specific DM scenarios such as that of the Higgsino DM model~\cite{Cirelli:2005uq}.
%This method has a major potential application in testing mixed-channel hypotheses, where DM is expected to annihilate or decay into a combination of leptonic and hadronic final states \mdc{The next is a repetition?}, as well as in refining predictions for DM models with multiple decay or annihilation pathways. Ultimately, our approach offers a flexible and generalized tool  to cast indirect-search DM limits on various DM particle models, thereby enhancing the interpretive power of $\gamma$-ray observations in the quest to uncover the nature of DM. 

The paper is structured as follows.  In \autoref{sec:dm_signal_model} we provide an introduction to the connection between experimental limits and the DM signal model. In \autoref{sec:methodology} we define the methodology of our algorithm. The results on real data are reported in  \autoref{sec:results} for various cases. We discuss and conclude in \autoref{sec:conclusion} and defer the more technical computation and further results in some appendices.

\section{DM signal model and experimental limits}
\label{sec:dm_signal_model}
The differential gamma-ray flux per unit energy $dE$ produced by the \textit{annihilation} of DM particles $\chi$ is given by:
\begin{equation}\label{eq:dm_flux_th_ann}
\frac{d\Phi}{dE}(E) = J_{\rm ann} \cdot \left(\frac{\langle \sigma v \rangle}{8\pi\;k\; m_{\chi}^2}\frac{dN_\gamma}{dE}\right),
\end{equation}
where the flux depends on two primary factors.
The \textit{astrophysical factor} $J_{\rm ann}$, defined as:
\begin{equation}\label{eq:j_ann}
J_{\rm ann} \equiv \int_{\Delta\Omega} d\Omega \int_{l.o.s.} dl\,\rho^2_{\chi}(l,\theta),    
\end{equation}
    accounts for the DM density distribution $\rho_{\chi}$ integrated along the line of sight $l$ and over the solid angle $\Delta\Omega$.  
The \textit{particle physics factor} (the term in parentheses) encodes the microscopic properties of the DM particle $\chi$. This factor includes the particle mass $m_{\chi}$, the average velocity-weighted annihilation cross section $\langle\sigma v\rangle$ and the differential photon yield per annihilation $dN_\gamma/dE$, determined by the specific particle-physics model considered. In   \autoref{eq:dm_flux_th_ann}  $k=1$ for a Majorana DM particle  and $k=2$ for a Dirac particle. In the following, we assume the $k=1$ and $\sigma=\langle\sigma v\rangle$ to simplify notation. 

A similar equation is obtained for the case of decaying DM but to improve readability we defer all decaying DM related discussion to \autoref{subsec:recast_decay}.

\begin{figure*}[h!t]
\centering
\includegraphics[width=0.95\linewidth]{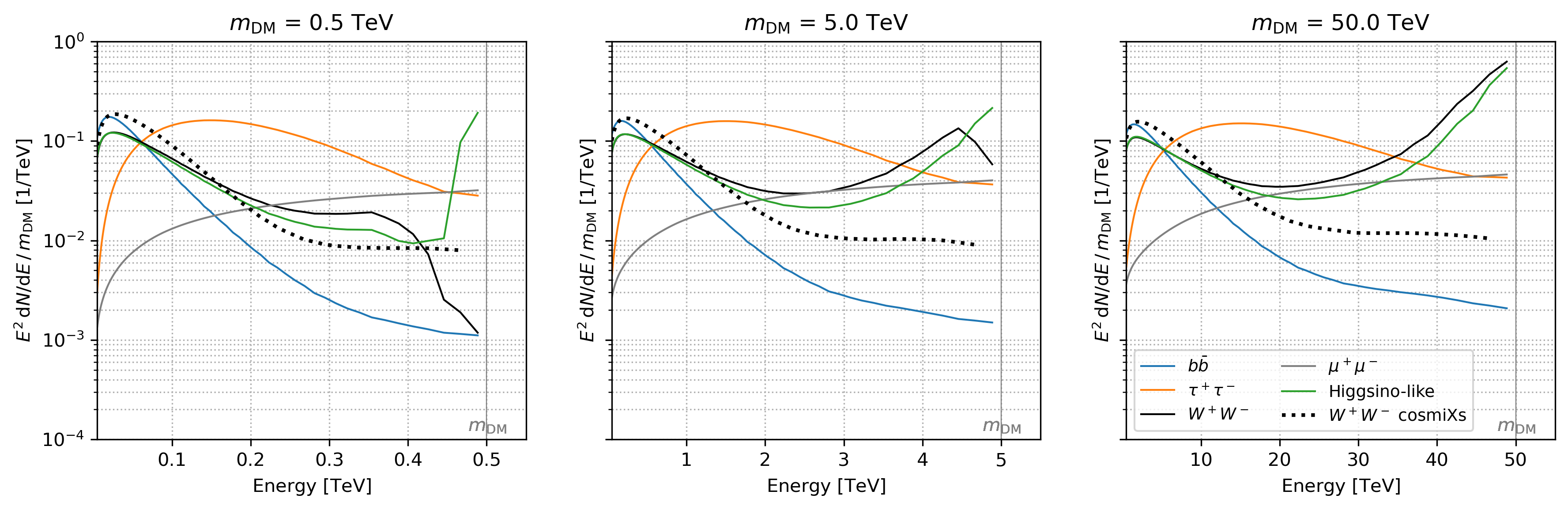}
\caption{DM $\gamma$-ray spectra for pure WIMP annihilation into specific channels, obtained with \texttt{gammapy}~\citep{gammapy:2023,gammapy:zenodo-1.2} and based on the PPPC parametrization (solid lines) by \citet{Cirelli:2010xx} for three values of masses and from \citet{arina2024cosmixs} (dashed line) for the $W^+W^-$ channel only. Also shown is an \textit{Higgsino-like} spectrum with annihilation into $W^+W^-,ZZ,\gamma\gamma/\gamma Z$ with branching ratios $BR_i=0.611,0.382,0.008$ respectively.}
\label{fig:dm_spectra}
\end{figure*}

\medskip
In either case, the term of the differential photon flux reads more precisely as:
\begin{equation}\label{eq:dnde}
\frac{dN_\gamma}{dE}=\sum_i\mathrm{BR}_i\frac{dN^i_\gamma}{dE}
\end{equation}
where the branching ratio terms $\mathrm{BR}_i$ encompass the specific microscopic nature of the DM and are computed for all possible annihilation channels $i=b\bar{b},\tau^+\tau^-,W^+W^-\ldots$ In practice, it is common for an indirect detection experiment to assume benchmark annihilation modes, e.g. set selectively $\mathrm{BR}_i=1$ for one channel only. This has the advantage that limits are more general and the disadvantage that they are not specific of an actual DM candidate. \autoref{fig:dm_spectra} shows some differential photon spectra of DM annihilation for three DM mass: $0.5, 5, 50$~TeV. The $b\bar{b}$ and $\tau^+\tau^-$ channels are often selected as representative examples of theoretical DM photon yields, the former being the softest with a signal peak at $m_{\mathrm{DM}}/20$ and the latter being the hardest with a peak at $m_\mathrm{DM}/3$. 
%In the case of DM decay, the major difference is that the cut-off happens at $m_\mathrm{DM}/2$ -- while preserving the spectral shape. 

On the other hand, some specific DM models are of interest, such as those of Higgsino~\cite{Cirelli:2005uq}. The thermal Higgsino, SUSY partner of the Standard Model (SM) Higgs, is the Lightest SUSY Particle, therefore a plausible thermal DM candidate, and would have a specific mass $m_h=1.08\pm0.02$~TeV~\citep{Bottaro:2022one} (which we approximate to $m_h=1.1$~TeV hereafter for simplicity). This mass is required to match the DM relic density. Higgsino annihilations would generate a prompt gamma ray continuum due to annihilation into $W^+W^-,ZZ$ as well as line-like signal to annihilation into $\gamma\gamma,\gamma Z$ with branching ratios $BR_i=0.611,0.382,0.008$ respectively. Further contributions for the Sommerfeld effect are expected~\citep{Rodd:2024qsi}. An Higgsino gamma-ray spectrum is also reported in \autoref{fig:dm_spectra}.

In \autoref{eq:dnde}, the photon yield is computed employing general-purpose Monte Carlo event generators for simulating high-energy particle collisions such as \texttt{Pythia}~\citep{Pythia8}. A standard has become the use of the so-called PPPC4-DM-ID "Poor Particle Physicist Cookbook for Dark Matter Indirect Detection"
from \citet{Cirelli2011PPPC}. We call this PPPC hereafter. More recently \citet{arina2024cosmixs} revisited the computation of source spectra from dark matter (DM) annihilation and decay by employing the \texttt{Vincia} shower algorithm within \texttt{Pythia}~\citep{Vincia}, an alternative parton shower plugin incorporating QED and QCD final state radiation as well as electroweak (EW) corrections with massive bosons, which were absent in the default \texttt{Pythia} shower. They included spin correlations and off-shell effects throughout the EW shower and tuned both \texttt{Vincia} and \texttt{Pythia} parameters to LEP data on pion, photon, and hyperon production at the Z pole~\citep{Fischer2016}. \citet{arina2024cosmixs} called this simulation \textit{cosmiXs}. The differences between PPPC4 and cosmiXs are shown for the $W^+W^-$ channel in \autoref{fig:dm_spectra}.
The discrepancy between PPPC4DMID and \texttt{cosmiXs} predictions, which increases 
towards higher dark matter masses, originates from the inclusion of electroweak (EW) 
corrections in PPPC4DMID. These corrections generate bump-like spectral features close 
to the kinematic endpoint $E \simeq m_\chi$, and their relative importance grows with 
increasing $m_\chi$. In addition, purely leptonic channels such as 
$\mu^+\mu^-$ exhibit sharp features due to final state radiation (FSR), which are also 
captured in PPPC4DMID. This explains the differences between the two predictions shown 
in \autoref{fig:dm_spectra} and motivates displaying both sets of spectra.

\subsection{The role of the instrument response}
From an experimental perspective, the expected number of gamma-ray events \( s_i \) within a given energy bin \( \Delta E'_i \) from an annihilating DM target with astrophysical factor $J_{\rm ann}=J$ observed for a time $T_{\mathrm{obs}}$, following \autoref{eq:dm_flux_th_ann}, is computed as:
\begin{align}
    \label{eq:k_i}
  & s_i = \langle  \sigma v \rangle \cdot K_i \quad \text{with}  \\ 
  \quad & K_i \equiv  \int_{\Delta E'_i} dE' \int dE \, A_{\gamma,\mathrm{eff}}(E) \cdot \mathcal{G}(E, E') \cdot \frac{dN_{\gamma}}{dE} \cdot \frac{T_{\mathrm{obs}}\, J}{8 \pi m_{\chi}^2}. \nonumber 
\end{align} 
where the instrumental response is characterized by an effective area \( A_{\gamma,\mathrm{eff}}(E) \) and an energy resolution \( \mathcal{G}(E,E') \) (where \( E' \) is the reconstructed photon energy). The corresponding formula for decaying DM is shown in \autoref{subsec:recast_decay}.  \autoref{eq:k_i} represents a simplified form that is more suitable for IACTs.  In more general cases, the instrument response functions depend not only on energy but also explicitly on spatial coordinates and on time. The complete expressions that account for these dependencies can be found, for example, in Eq.~(16) and Eq.~(17) of Ref.~\cite{rico2020gamma}.

It is important for the purpose of this work to discuss the two terms in more detail. The effective area is the product of the instrument 'aperture' $A_\gamma(E)$ times the detection efficiency $\epsilon_\gamma(E)$:
\[
A_{\gamma,\mathrm{eff}}(E)=A_\gamma(E)\cdot\epsilon_\gamma(E)\quad\mathrm{with}\quad\epsilon_\gamma=\frac{N^{\mathrm{rec}}_\gamma(E)}{N^{\mathrm{sim}}_\gamma(E)}
\]
where $N^{\mathrm{sim}}_\gamma(E)$ is the number of simulated signal events at energy $E$ and $N^{\mathrm{rec}}_\gamma(E)$ is the number of events actually observed at that energy.  Note that a different acceptance $\epsilon_b(E)$ can be defined for background events. 
In IACTs, the detection efficiency also 
depends on the impact distance $r$ (and on the offset). One may write
\begin{equation}
A_{\gamma,\mathrm{eff}} = \int \mathrm{d}r \; p(r) \, A_{\gamma}(E,r) \, \epsilon_\gamma(E,r) \, ,
\end{equation}
where $p(r)$ is the distribution of impact distances, $A(E,r)$ the geometric acceptance, 
and $\epsilon_\gamma(E,r)$ the photon detection efficiency. In practice, collaboration--provided 
$A_{\mathrm{eff}}(E)$ curves are already folded over the relevant distributions for each 
observation configuration.

\begin{figure*}[h!t]
\centering
\includegraphics[width=0.35\linewidth]{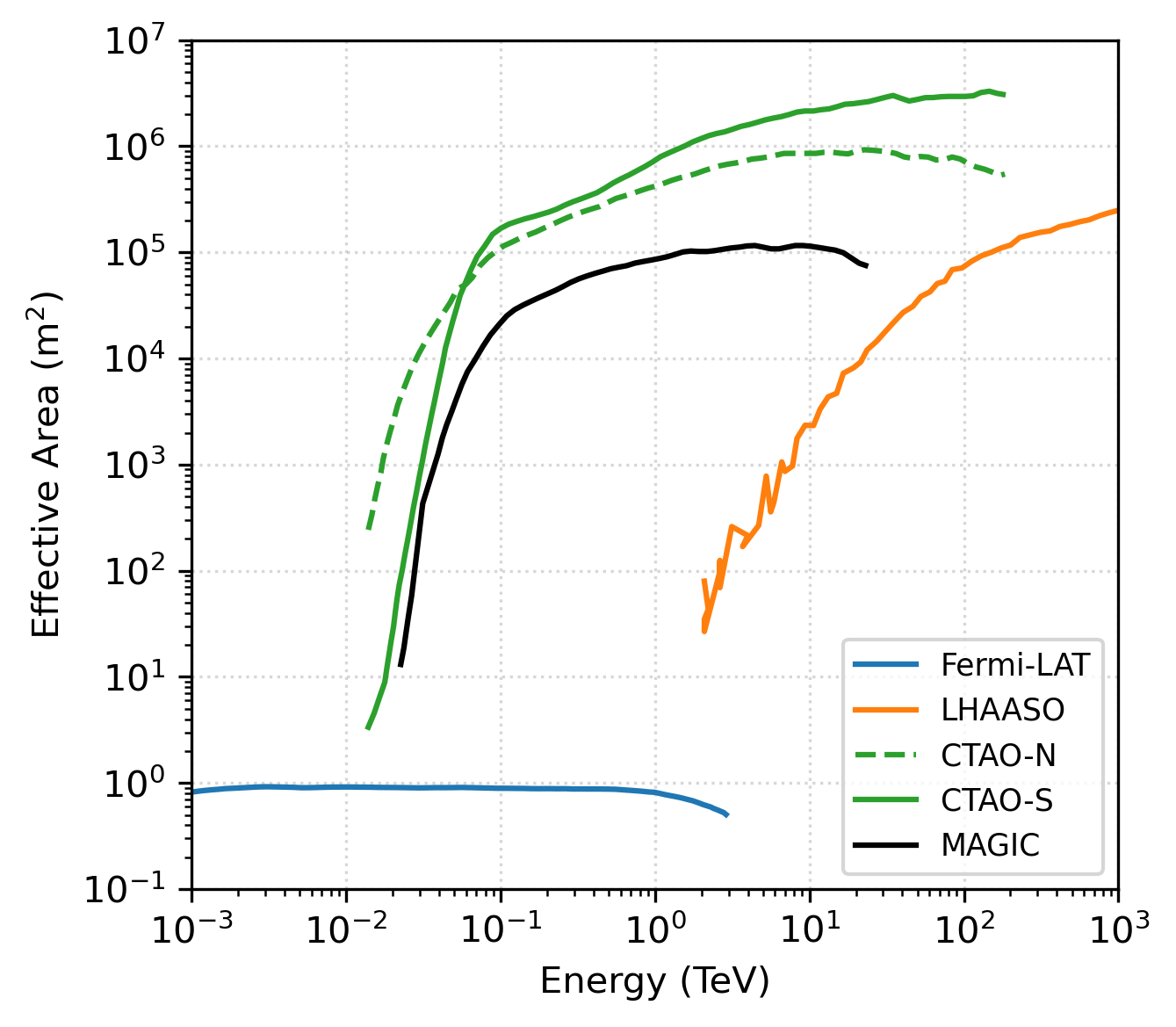}
\includegraphics[width=0.35\linewidth]{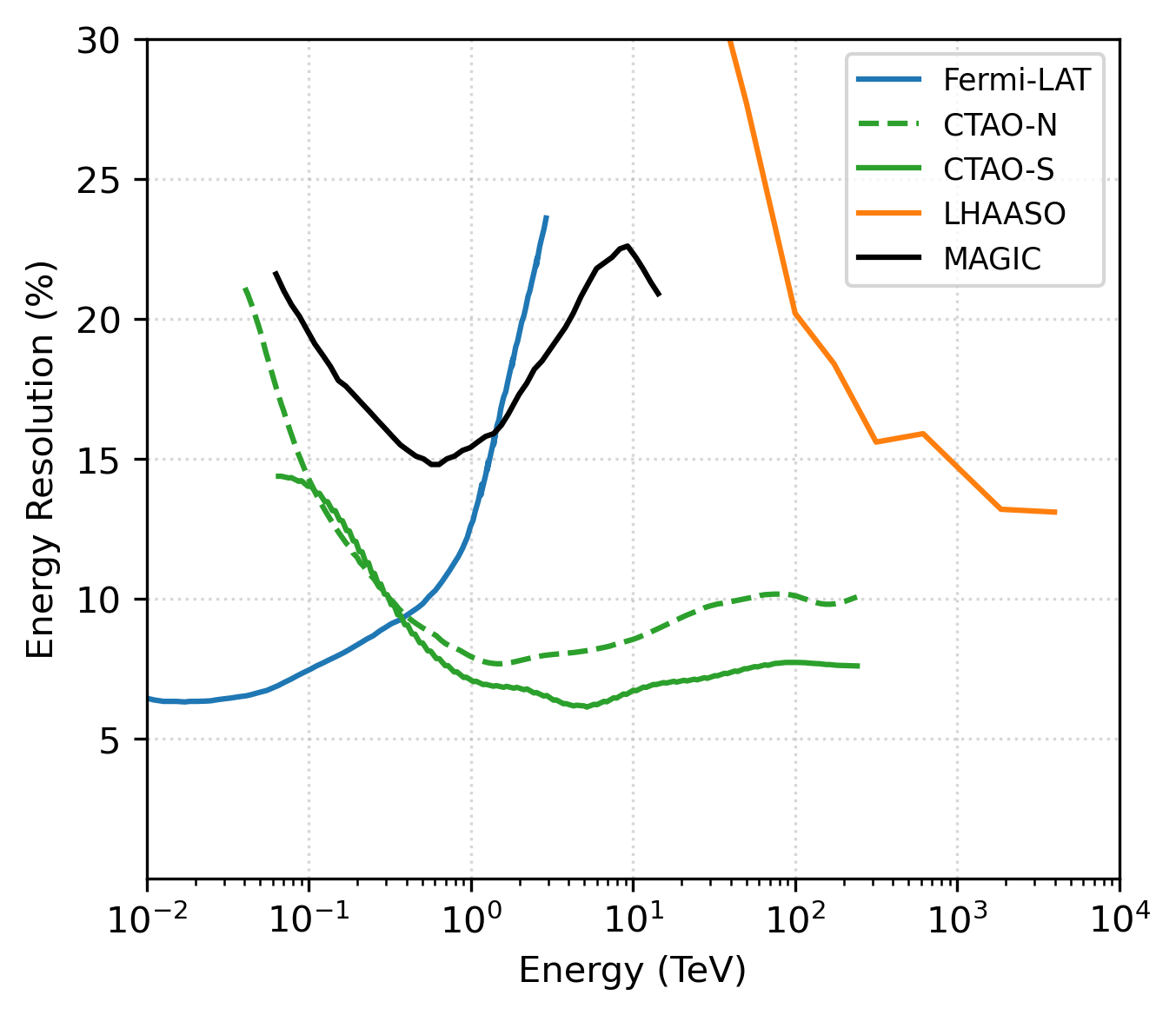}
\caption{Effective area (left) and energy resolution (right) for  \textit{Fermi}-LAT~\citep{aeff_FermiLAT}, CTAO~\citep{CTAOwebsite}, MAGIC~\citep{MAGIC:performance}, and LHAASO \citep{LHAASO:2019qtb}.} 
\label{fig:aeff_enres}
\end{figure*}

Some examples of effective areas are shown for MAGIC, CTAO, \textit{Fermi}-LAT and LHAASO in \autoref{fig:aeff_enres} (left panel). The energy resolution is well described in terms of a Gaussian distribution centered around the true energy value $E$ and with width proportional to the energy. This is true for all gamma-ray instruments described above. The energy resolution is roughly $5-20\%$ for \textit{Fermi}-LAT and IACTs depending on the energy, and somewhat larger for SFDs, as shown in \autoref{fig:aeff_enres} (right panel).  

The energy resolution and effective area change observation-by-observation and depend both on the actual instrument status during observation (e.g. mirror reflectivity for IACTs, hardware settings, etc.), observational conditions (e.g. the zenith angle of the target) or external factors such as the amount of diffuse light of the night sky (for IACTs) or albedo light (for \textit{Fermi}-LAT) and so on. Therefore, there is no unique effective area, and in data reconstruction, the energy dispersion and effective area are computed \textit{for a specific data analysis}\footnote{Many analyses now compute run-by-run  IRFs~\cite{holler2020run} to incorporate detector/atmosphere/zenith/offset variations; this is also the envisioned standard for CTAO.}. These are only rarely reported in published results, and externals relies on generic benchmark effective area published elsewhere by the collaborations, a fact we will discuss at length below.

\medskip
In searches for DM signals within gamma-ray data, no conclusive indirect detection has yet been reported\footnote{Possible exceptions, such as the gamma-ray excess from the Galactic Center observed by \textit{Fermi}-LAT, remain controversial and inconclusive~\cite{ackermann2017fermi}.}, implying that current observations are   consistent with \( s_i = 0 \) in \autoref{eq:k_i}, and consequently, with \( \langle \sigma v \rangle = 0 \). As a result, analyses have focused on computing upper limits mainly in the DM annihilation cross section \( \langle \sigma v \rangle \) (or \( dN_\gamma\langle \sigma v \rangle/dE\) in \cite{Rodd:2024qsi}).  
These upper limits depend primarily on two key factors: (i) the theoretical prediction of the expected number of signal events, given by \autoref{eq:k_i} and (ii) the experimental gamma-ray excess data counts $n_{exc}$, which is a possible realization of the former. 

However, published results are typically only expressed in terms of $\langle \sigma v \rangle$-limits  with no report of the actual observed gamma-ray excesses or the actual instrument response functions for that specific analysis. As we show in the next section, the publication of this supplementary information would allow any experimental limit to be flexibly reinterpreted (``recast'') to derive constraints on alternative DM models without accessing the original data and likelihood analysis. 
%Furthermore, the publication of the ancillary information would alone allow to "forecast" limits on any parametric DM model. 
An example of the criticality of missing experimental information has already been studied in detail in \citet{Rodd:2024qsi}, who, for example, found an inconsistency in the estimation of the effective area related to published IACT results, which cannot be verified a posteriori. We see that this omission impedes the reproducibility of the full experimental results.

\section{Methodology to cast and forecast $\langle \sigma v \rangle$}
\label{sec:methodology}
In this section, we outline the methodology of the algorithm used to forecast and recast the experimental limits of DM signals. As the formalism presented here is sufficiently generic to be adapted to other potential scenarios, such as the decaying DM, we restrict our discussion to the annihilation case and defer the decaying DM case to \autoref{subsec:recast_decay}.

\subsection{Forecasting and recasting upper limits on \sv\ }
\label{sec:method_exact}

%To determine the upper limit (UL) on the DM annihilation cross section, \( \langle \sigma v \rangle^{UL} \), a binned likelihood analysis 
%is typically performed~\cite{rico2020gamma}. Denoting from now on, for simplicity, our parameter of interest, the averaged cross section \( \langle \sigma v \rangle \) as \( \sigma \), the total log-likelihood  can be written as~\cite{DAmico:2022psx}:

To determine the upper limit (UL) on the DM annihilation cross section, 
$\langle \sigma v \rangle^{\mathrm{UL}}$, a binned likelihood analysis 
is typically performed (see, e.g.,~\cite{rico2020gamma,abramowski2011search,lefranc2016dark}). 
Denoting, for simplicity, our parameter of interest, the averaged cross section 
$\langle \sigma v \rangle$, as $\sigma$, the total log-likelihood can be written 
as in~\cite{DAmico:2022psx,conrad2015statistical}. 
\begin{equation}
    -2 \ln\mathcal{L}(\sigma) 
    = - 2 \sum_i \ln \mathcal{L}_i (s_i(\sigma) \; | \; D_i ) 
    \equiv  2 \sum_i f_i(s_i(\sigma)),
    \label{Eq:general_likelihood}
\end{equation}
where the index \( i \) spans through bins, typically energy bins. Within each bin, \( D_i \) represents the observed data (such as the counts in each bin)
and \( s_i \) are the expected signal counts computed from \autoref{eq:k_i}. 
It is worth recalling that both IACT and \textit{Fermi}-LAT analyses 
routinely make use of spectral and spatial binned likelihoods. 
For simplicity and clarity, in the present work we restrict ourselves 
to the spectral (energy-binned) case only.

\noindent
Defining \( f' \) and \( f'' \) as the first and second derivatives of \( f(s) \) with respect to \( s \), 
%\begin{align}
 %   f''_i(s_i) &= \ldots\\
 %    f''_i(s_i) &= 
%    \ldots 
 %        \label{eq:derivative_f}
%\end{align}
and using   \autoref{eq:k_i}, the log-likelihood can be expanded \cite{wald1943tests} around the value \(\hat{\sigma}\) that minimizes \autoref{Eq:general_likelihood} as \footnote{This is because the first-order derivative \(\sum_i f_i'\left(s_i(\hat{\sigma})\right) = 0\) at \(\hat{\sigma}\), while the zero-order term of the Taylor expansion vanishes if we assume a properly profiled likelihood, i.e., \( \sum_i f_i \left(s_i(\hat{\sigma})\right) = 0 \). See Ref. \cite{wald1943tests} for a detailed discussion on the approximation of the likelihood ratio for a single parameter of interest.}
\begin{equation}\label{Eq:likelihood_Taylor2}
    -2 \ln \mathcal{L}(\sigma)  \simeq \sum_i K_i^2\, f_i''(K_i \hat{\sigma})\,(\sigma - \hat{\sigma})^2.
\end{equation}

Finally \(\sigma^{UL}\) can be obtained by setting \(-2 \ln \mathcal{L} = \lambda\) in \autoref{Eq:likelihood_Taylor2}, where \(\lambda\) determines the confidence level (CL) used for the UL, giving~\footnote{For instance, a one-sided \(95\%\) confidence level  UL is obtained with \(\lambda = 2.71\)~\cite{Rolke:2004mj}. To compute a lower limit instead of an upper limit, the ``+'' sign in   \autoref{ UpperLimit} should be replaced with a ``-'' sign.}
\begin{equation}
     \sigma^{UL} \simeq \hat{\sigma} + \sqrt{\frac{\lambda}{\sum_i K_i^2 \; f_i''(K_i \hat{\sigma})}}.
     \label{ UpperLimit}
\end{equation}

Defining \(n_i\) as the counts observed in the \(i\)-th bin,  the functions \(f_i\)  take the form~\cite{DAmico:2022psx} :
\begin{align}
    f_i(s_i) &= s_i - n_i \ln(s_i + b_i) + C,  \quad \rm{or} \label{eq:cash_f}\\
     f_i(s_i) &= 
    s_i  - n_i \ln \left( s_i +  b_i \right) \nonumber \\ 
    & +  (1 + \alpha)\, b_i - 
    m_i \ln \left( \alpha b_i \right) 
     + C,
         \label{Eq:Lklratio_b2}
\end{align}
where \(C\) is a constant independent of the model parameters. The former formulation is known in the literature as the Cash statistic \citep[C-statistic,][]{Cash:1979vz}, which is applied in Poisson processes when the expected background counts \(b_i\) are known for each bin \(i\). 
The latter expression applies when $b_i$  are estimated from a background control region (referred to as the OFF region), which provides background counts \(m_i\) for each bin \(i\), with \(\alpha\) a scaling factor accounting for differences in exposure or efficiency between the signal (ON) and background (OFF) regions~\footnote{$\alpha =\frac{\rm{Exposure}_{\rm{OFF}}}{\rm{Exposure}_{\rm{ON}}}$. In \autoref{Eq:Lklratio_b2} we assume no signal leak in the background region.}. 

It can be shown \cite{cowan2011asymptotic} that the second derivatives \( f_i'' \) are linear with respect to the data values \( n_i \) and/or \( m_i \) (see    \autoref{App:Second_Der} for the derivation of $f''$ applied to \autoref{eq:cash_f} and \autoref{Eq:Lklratio_b2} ). Consequently, the expectation value of \( f_i'' \) can be calculated exactly using the expected values of the data\footnote{These expected values are often referred to as the Asimov dataset \cite{cowan2011asymptotic}.}. Under the null hypothesis of no DM signal, these expected values are obtained by putting $\hat{\sigma} = 0$ and \autoref{ UpperLimit} simplifies to:
\begin{align}
    \sigma^{UL}  \simeq \sqrt{\frac{\lambda}{\sum_i K_i^2/b_i}} \qquad & \text{[Cash]} \\
    \sigma^{UL}  \simeq \sqrt{\frac{\lambda}{\sum_i K_i^2/ ((1+\alpha^{-1}) b_i)}} \qquad &\text{[ON/OFF]}
    \label{eq:UpperLimit_Asimov}
\end{align}
where the former equation is applied in the context of the C-statistic (\autoref{eq:cash_f}), and the latter applies when background counts are estimated from an OFF region (\autoref{Eq:Lklratio_b2}).  

\medskip\noindent
It is worth highlighting four important implications (the first two are well known and can be seen as consistency checks) of the result presented in \autoref{eq:UpperLimit_Asimov}:

\begin{itemize}
    \item \textbf{Impact of Background Knowledge:} If the expected background counts $ b_i$ in the ON region are not known precisely and must be estimated from an OFF region, the ULs are generally less stringent. This is because the factor \( 1+\alpha^{-1} \) is always greater than 1, which reduces the denominator in \autoref{eq:UpperLimit_Asimov}, leading to a larger value of the UL. This behavior is expected, as precise knowledge of the background improves the reliability and tightness of exclusion limits by reducing statistical uncertainty. Moreover, in the limit where the exposure in the OFF region becomes infinitely larger than that of the ON region, \( \alpha^{-1} \to 0 \), and the expression for the upper limit asymptotically approaches that of the Cash statistic. This is expected, as an infinitely large OFF exposure yields a nearly perfect estimate of the background, reducing the ON--OFF analysis to the standard Cash-statistic case with known background.\bigskip

    \item \textbf{Advantage of Multi-bin Analyses:} As demonstrated in    \autoref{App:Inequality}, the inequality
\begin{equation}
    \sqrt{\sum_i \frac{K_i^2}{b_i}} \geq \frac{\sum_i K_i}{\sqrt{\sum_i b_i}}
    \label{eq:inequality}
\end{equation}
holds in general. This implies that, for a fixed total exposure and background, performing the analysis using multiple energy bins leads to a larger denominator in \autoref{eq:UpperLimit_Asimov}, and consequently, to more stringent (i.e., lower) upper limits on the DM annihilation cross section. In other words, binning the data allows for a more efficient exploitation of the energy dependence of the expected signal, enhancing the sensitivity of the search.  \bigskip

    \item \textbf{Forecasting Upper Limits:} Once the instrument's response functions (IRFs)—namely the effective area, energy dispersion, and background model—are known, \autoref{eq:UpperLimit_Asimov} allows one to directly forecast the UL on the DM cross section under the null hypothesis of no DM signal\footnote{It is worth noting, however, that if the full IRF information is available, one could perform the likelihood analysis directly under the null hypothesis (i.e., assuming no signal). }. This provides a fast and accurate way to estimate experimental sensitivity, removing the need for Monte Carlo simulations\footnote{The IRFs and energy dispersion themselves are typically derived from dedicated MC production and/or calibration; here we mean that, once IRFs are available, \autoref{eq:UpperLimit_Asimov} gives the ULs directly under the no-signal hypothesis.} and likelihood analyses for each realization of the data. This is discussed and validated in \autoref{sec:results_forecast}. \bigskip

    \item \textbf{Recasting Across Models:} A practical consequence of \autoref{eq:UpperLimit_Asimov} is that, given two different DM models, \( M_0 \) and \( M_I \), each predicting a distinct gamma-ray energy spectrum \( dN^0_{\gamma}/dE \), and \( dN^I_{\gamma}/dE \), respectively,  the UL on the cross section for model \( M_I \), denoted as \( \sigma^{UL}_I \), can be derived from the UL for model \( M_0 \), denoted as \( \sigma^{UL}_0 \), via the relation:
    \begin{equation}\label{eq:ratio}
        \sigma^{UL}_I = \sqrt{ \frac{\sum_i K^2_{0,i} / b_i}{\sum_i K^2_{I,i} / b_i} } \cdot \sigma^{UL}_0,
    \end{equation}
    where \( K_{0,i} \) and \( K_{I,i} \) are computed using \autoref{eq:k_i}, with \( dN_{\gamma}/dE \) evaluated according to models \( M_0 \) and \( M_I \), respectively. This relation enables the recasting of ULs between models without repeating the full likelihood analysis, which would otherwise require access to the observed data and the statistical machinery used in the original limit computation. The expression \autoref{eq:ratio} is valid for both the Cash and ON/OFF case because in the latter the \( 1+\alpha^{-1} \) term of \autoref{Eq:Lklratio_b2} simplifies.

    This methodology is discussed and validated through MC simulations in \autoref{sec:mc}, and further tested against published upper limits from various experiments in \autoref{sec:results_cast_benchmarks} for different annihilation channels. Its versatility is further demonstrated in \autoref{subsec:recast_decay} by recasting annihilation limits to decay scenarios, and in \autoref{sec:results_cast_higgsino} by translating limits for specific DM models (e.g., Higgsino) and for different photon yield prescriptions.

\end{itemize}

\paragraph{Note on nuisance parameters}  
At this point, it is worth noting that the log-likelihood functions considered in \autoref{Eq:Lklratio_b2} do not explicitly include nuisance parameters beyond the expected background counts $b_i$. In dark matter searches, one of the most relevant nuisance parameters is the astrophysical $J$-factor \autoref{eq:j_ann}, which carries uncertainties from the modeling of the DM density distribution. Since the $J$-factor acts as a global ``strength'' parameter (that is, higher values of $J$ imply proportionally higher expected signal counts $s_i$), its inclusion as a nuisance parameter in the likelihood effectively rescales the resulting ULs on the annihilation cross section by a constant factor greater than one~\footnote{This rescaling argument applies to spectral-only likelihoods. In spatial and spectral analyses, the spatial morphology is tied to $J$ and complicates a pure rescaling.}. Consequently, the \textit{recasting procedure} presented in \autoref{eq:ratio} is unaffected by whether the $J$-factor is treated as fixed or as a nuisance parameter. For the \textit{forecasting procedure}, however, if the $J$-factor is treated as a nuisance parameter, one cannot use \autoref{eq:UpperLimit_Asimov} directly but should perform the full likelihood analysis. 

\subsection{Approximate algorithm when dealing with missing Instrument Response Functions}
\label{sec:method_approx}

In many cases, especially when analyzing published results from gamma-ray telescopes, the detailed instrumental information used to derive upper limits on the DM annihilation cross section is not made publicly available, as we argued before. As a result, crucial inputs such as effective area \( A_{\gamma,\mathrm{eff}}(E) \) and background modeling \( b_i \) cannot be accessed, preventing direct computation of \autoref{eq:k_i} and thus obstructing complete recasting of ULs onto alternative DM models via \autoref{eq:ratio}.

Nevertheless, an approximate recasting remains feasible by rewriting \autoref{eq:ratio}, using the definition of $K_i$ of \autoref{eq:k_i}, in the following form:

\begin{equation}
 \sigma^{UL}_I = \sqrt{ \frac{\sum_i K^2_{0,i}/b_i}{\sum_i K^2_{I,i}/b_i} } \cdot \sigma^{UL}_0 
 \equiv
 \sqrt{ \frac{\sum_i \left( V_i \cdot \Delta N^0_{\gamma,i} \right)^2 }{\sum_i \left( V_i \cdot \Delta N^I_{\gamma,i} \right)^2 } } \cdot \sigma^{UL}_0,
\label{Eq:recast_approx}
\end{equation}

where \( \Delta N_{\gamma,i} \equiv \int_{\Delta E'_i} dE \, \frac{dN_{\gamma}}{dE} \) is the intrinsic number of photons predicted by the DM model in the \( i \)-th energy bin, and the factor

\begin{equation}
    V_i \equiv \frac{K_i}{\sqrt{b_i} \cdot \Delta N_{\gamma,i}}
    \label{Eq:Vi_main_text}
\end{equation}

\noindent
encodes the instrument response and background scaling. 
As shown in \autoref{App:Vi}, under mild assumptions --- such as limited bin-to-bin energy spillover (as in the case of good energy resolution) --- the \( V_i \) coefficients can be treated as approximately model independent, that is, they solely depend on the IRFs rather than the specific DM spectral shape.

This property allows one to numerically infer the \( V_i \) values by requiring \autoref{Eq:recast_approx} to hold for two benchmark DM channels whose upper limits \( \sigma^{UL}_0 \) and \( \sigma^{UL}_I \) are publicly available. In practice, this means that by solving an optimization problem (e.g., least squares) over a set of energy bins, one can extract an effective \( V_i \) array that can then be used to recast the ULs onto any other DM model, even when the full IRF information is missing. This approach is validated with Monte Carlo simulations in \autoref{sec:mc}  and with real data from published results in \autoref{sec:results_cast_benchmarks}.

%Secondly, if the summation \(\sum K_i^2 / b_i\) is dominated by a single bin \(j\), that is, \(\sum K_i^2 / b_i \approx K_j^2 / b_j \equiv K^2 / b\), or if the analysis is conducted using only a single bin, then the UL expression simplifies to
%\begin{equation}
%    \sigma^{UL} \simeq \frac{\sqrt{\lambda \; b}}{K}
%    \label{eq:UpperLimit_Asimov_2}
%\end{equation}
%Therefore, in this case, the upper limit is decoupled into two distinct components. The first component depends solely on the signal model and the telescope’s IRF, as encapsulated by $K$ in \autoref{eq:k_i}. The second component depends exclusively on the expected background counts $b$.

\subsection{Test on Toy MC simulatiuons}
\label{sec:mc}

\begin{figure*}[t]
    \centering
    \includegraphics[width=0.75\linewidth]{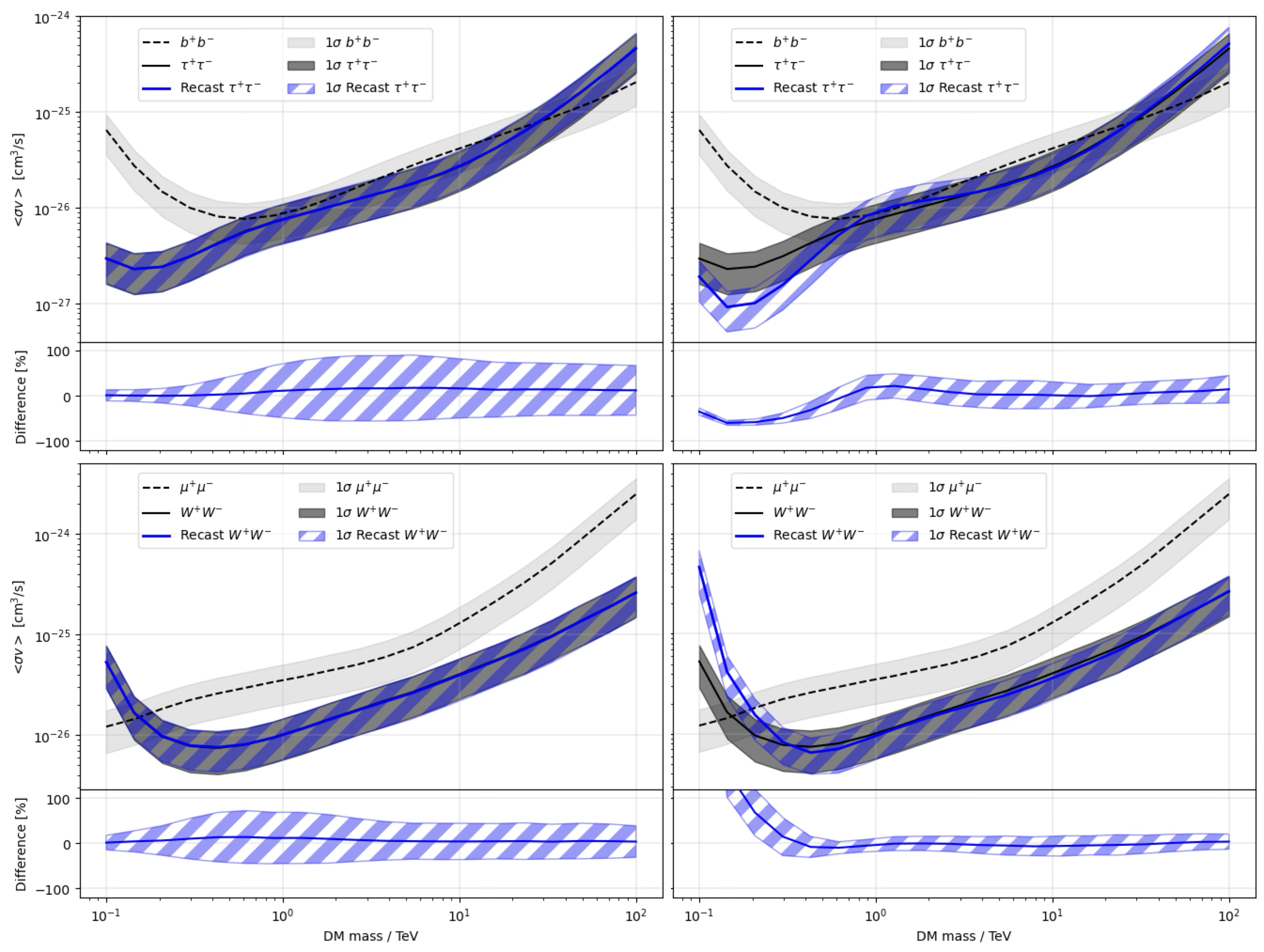}
  \caption{
Each panel summarizes $10^{5}$ Monte Carlo realizations generated under the null (no-signal) hypothesis with the CTAO IRFs.  
\emph{Left column:} recast performed with the exact IRF using \autoref{eq:ratio}.  
\emph{Right column:} recast performed with the IRF–free approximation of \autoref{Eq:recast_approx}.  
\emph{Top row:} ULs for the \(\tau^{+}\tau^{-}\)  channel reconstructed from \(b\bar{b}\)  ULs.  
\emph{Bottom row:} ULs for the \(W^{+}W^{-}\) channel reconstructed from \(\mu^{+}\mu^{-}\) ULs.  
Color code: light-grey bands show the \(1\sigma\) containment of the benchmark (input) channels; dark-grey bands show the \(1\sigma\) containment of the \emph{true} ULs for the target channel; zebra pattern (alternating blue and transparent stripes) show the corresponding recast ULs.  
Black curves give the mean ULs over all simulations, while blue curves give the mean of the recast ULs.  
The lower sub-panel in each plot displays the fractional difference in percentage between true and recast ULs.}
    \label{fig:Toy_MC_figure}
   \end{figure*}

To test the accuracy of the algorithms in \autoref{eq:ratio} and \autoref{Eq:recast_approx}, we generated \(10^{5}\) toy Monte Carlo (MC) realizations under the null hypothesis of no DM signal. For each realization, we draw Poisson distributed counts \(n_i\) (ON region) and \(m_i\) (OFF region) in every energy bin.  
      The publicly available CTAO instrument response functions (IRFs) were adopted for the effective area and energy dispersion~\cite{ctao_irfs}.  
      The values used for the observation time \(T_{\mathrm{obs}}\) and $J$-factor in \autoref{eq:k_i} do not affect the test, because they appear in every \(K_i\) and thus cancel in the ratio of \autoref{eq:ratio}. Using the binned likelihood of \autoref{Eq:Lklratio_b2}, we derived \(\sigma^\mathrm{UL}\) for each DM mass \(m_\chi\) and for four annihilation channels: \(\tau^+\tau^-\), \(b\bar{b}\), \(\mu^+\mu^-\), and \(W^+W^-\).  
      Photon spectra \(dN_\gamma/dE\) were taken from \texttt{PPPC4DMID} tables via \texttt{gammapy} \cite{gammapy:2023,gammapy:zenodo-1.2}. For each generated MC, we then  performed two validations:

\begin{enumerate}
\item \textbf{Exact recast (IRF known).}  
      The ULs for \(\tau^+\tau^-\) and \(W^+W^-\) were recast from those of  \(b\bar{b}\)  and \(\mu^+\mu^-\), respectively, using \autoref{eq:ratio}.  
      The results are shown in the left panels of \autoref{fig:Toy_MC_figure}.  \medskip

\item \textbf{Approximate recast (IRF unknown).}  
      A second set of recasts was performed with no knowledge of the true IRF, replacing \autoref{eq:ratio} by the approximation in \autoref{Eq:recast_approx}.  
      The factors \(V_i\) were inferred by requiring \autoref{Eq:recast_approx} to hold between two benchmark channels (\(b\bar{b}\ \, , \;  W^+W^- \) and \(\mu^+\mu^- \, , \; b\bar{b} \), respectively).  
      The corresponding results appear in the right panels of \autoref{fig:Toy_MC_figure}.  \medskip
\end{enumerate}

\autoref{fig:Toy_MC_figure} displays, for each channel, the \(1\sigma\) containment band of the ULs versus \(m_\chi\) (light gray for the benchmark channels, dark gray for the \emph{true} ULs of the target channels, and blue-transparent zebra stripes for the recast ULs).  
The black solid (dashed) lines indicate the mean ULs for the \emph{true}  (benchmark) simulations; while the solid blue lines show the recast means.  
The same color coding is used in the lower sub-panels which present the fractional difference in percentage between the true and recast ULs.

When the exact IRF is used, the recast ULs (zebra pattern) overlap the true ULs (dark grey) within statistical fluctuations across the entire mass range. The corresponding mean curves -- black for the true limits and blue for the recast ones -- are indistinguishable across the full mass range.
With the IRF-free approximation, good agreement is recovered for \(m_\chi\gtrsim 200~\mathrm{GeV}\); at lower masses the mismatch grows because too few mass bins contribute to a reliable determination of \(V_i\). 
Outside of this low-mass regime, the approximate method reproduces the true limits within \(1\sigma\) uncertainty, confirming its utility when only published ULs are available.

Lastly, the lower subpanels in each plot show that the IRF-free approach yields a visibly smaller relative difference between the true and recast ULs.  
This is because the coefficients \(V_i\) are fitted in every Monte Carlo realization using the ULs from the benchmark channels that themselves carry Poisson fluctuations; thus the fit absorbs part of that noise, so true and recast ULs inherit the same realization-dependent fluctuations, and their relative difference is less dispersed.  
In contrast, when the exact IRF is used directly in \autoref{eq:ratio},  Poisson noise is not taken into account, producing a greater spread between the true and recast ULs.

\section{Results with real data}
\label{sec:results}

\subsection{Forecasting Upper Limits with CTAO IRFs}\label{sec:results_forecast}

We now validate the forecasting formula presented in \autoref{eq:UpperLimit_Asimov} by computing ULs on the DM annihilation cross section for the dwarf spheroidal galaxy Draco I, as presented in Ref.~\cite{CTAO-dphs}. To perform this test, we adopt the same observational assumptions as in Ref.~\cite{CTAO-dphs}: a $J$-factor of \(10^{18.7} \, \text{GeV}^2/\text{cm}^5\) integrated over a cone of radius \(0.5^\circ\), and a total observation time of 100 hours. The publicly available instrument response functions (IRFs) of the Cherenkov Telescope Array Observatory (CTAO)~\cite{ctao_irfs} are used.

\begin{figure*}[h!t]
    \centering
    \includegraphics[width=0.7\linewidth]{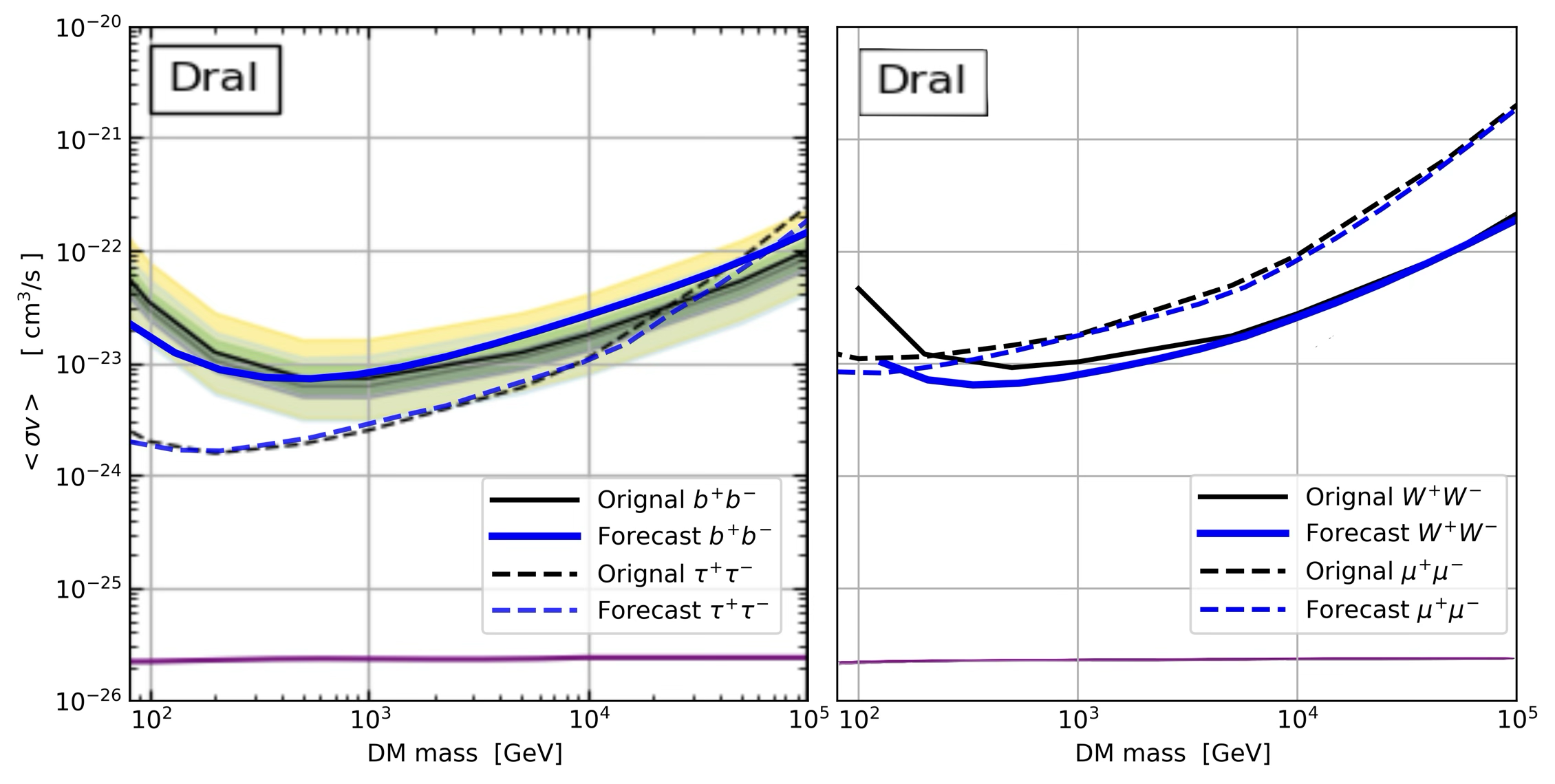}
    \caption{
    Forecasted 95\% CL upper limits on the DM annihilation cross section toward Draco I, assuming a $J$-factor of \(10^{18.7} \, \text{GeV}^2/\text{cm}^5\) within \(0.5^\circ\), 100 hours of observation time and the CTAO IRF ~\cite{ctao_irfs}. 
    Left: Comparison between forecasted ULs (blue lines) and published CTAO results (black lines and bands) for the \(b\bar{b}\) and \(\tau^+\tau^-\) channels. Our forecast ULs (blue solid and dashed lines) are overlaid on the original CTAO plot, figure 8 of Ref. ~\cite{CTAO-dphs}.
    Right: Same comparison for the \(W^+W^-\) and \(\mu^+\mu^-\) channels. 
    }
    \label{fig:forecast_ctao}
\end{figure*}

The left panel of Fig.~\ref{fig:forecast_ctao} shows the comparison between our forecast ULs and those published in Ref.~\cite{CTAO-dphs} for the \(b\bar{b}\) and \(\tau^+\tau^-\) annihilation channels. The right panel displays the same comparison for the \(W^+W^-\) and \(\mu^+\mu^-\) channels. In the left panel, the forecast ULs (blue solid and dashed lines) are overlaid on the CTAO results (black solid and dashed lines with uncertainty bands) presented in figure 8 of Ref. ~\cite{CTAO-dphs}.

We observe good agreement between our forecast and the published CTAO limits 
across all tested channels at high DM masses. At lower masses, discrepancies 
become more visible. To quantify this effect, we computed the relative 
difference between our recast limits and the published mean expected values. 
For the $b\bar{b}$ and $W^+W^-$ channels, the discrepancy reaches up to 
$\sim 70\%$ at the lowest tested masses and decreases rapidly to the 
few--percent level above $\sim 1$~TeV. For the $\tau^+\tau^-$  and $\mu^+\mu^-$ channel, the 
variation remains at the $\sim 30\%$ level across the mass range, typically improving 
to the $5$--$15\%$ level at higher masses.

This validates the reliability of \autoref{eq:UpperLimit_Asimov} as a practical and accurate tool for computing sensitivity projections from publicly available IRFs, without the need for full-scale Monte Carlo simulations and likelihood analysis.

\subsection{Casting of benchmark annihilation channels}
\label{sec:results_cast_benchmarks}

Since published results typically lack access to instrument response functions (IRFs) and expected background counts, we first validate the approximate recasting procedure of \autoref{Eq:recast_approx} by translating the published upper limits from one annihilation channel to another.

\begin{figure*}[h!t]
%\begin{adjustwidth}{-\extralength}{0cm}
    \centering
    \includegraphics[width=0.75\linewidth]{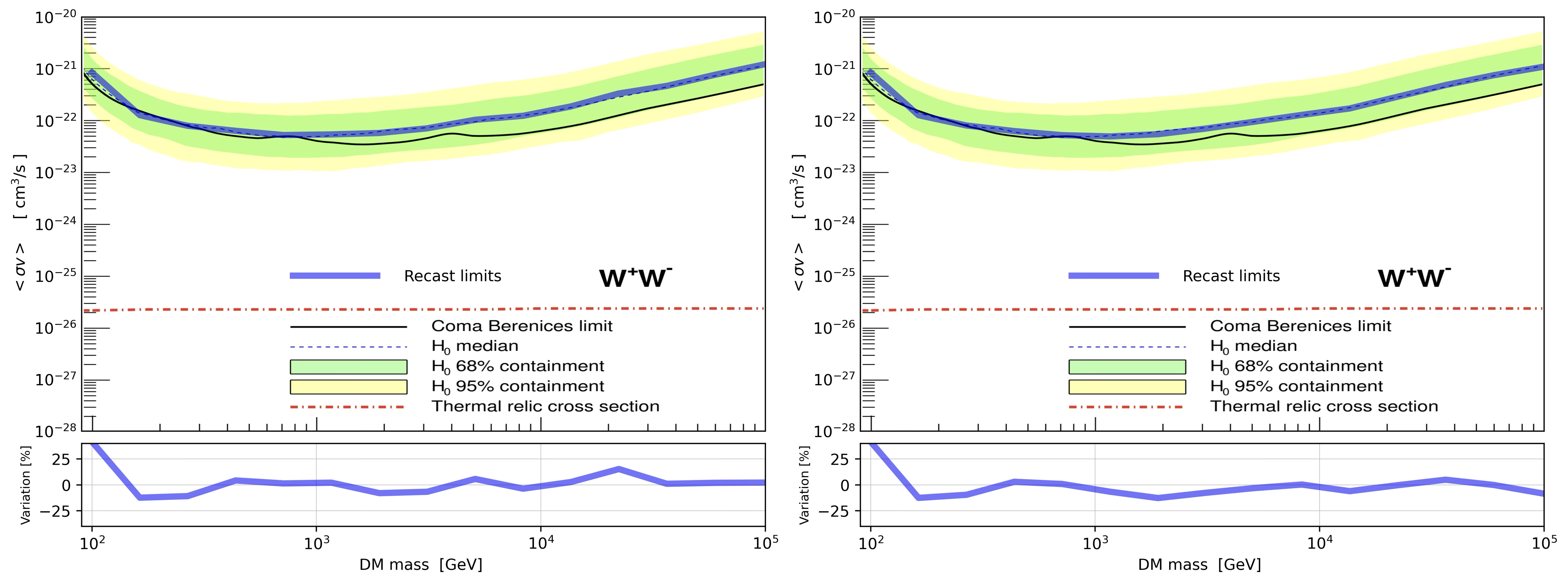}
    \caption{
In both panels  the image represents the originally published  ULs from the MAGIC collaboration~\citep{MAGIC:2021mog}. The black line denotes the UL (solid for the observed one, dashed for the median), the green/yellow shaded regions denote the $68\%$/$95\%$ band of the null hypothesis,  as derived from Monte Carlo sampling.
Overlaid is our recast UL (thick solid blue line) obtained using \autoref{Eq:recast_approx} to recast limits for \(W^{+}W^{-}\) annihilation from the \(b \bar{b}\) channel. \textit{Left:} ULs obtained using coefficients \(V_i\) inferred using the \(\tau^{+}\tau^{-}\) channel as a second benchmark.  
\textit{Right:} ULs obtained using coefficients \(V_i\) inferred using the \(\mu^{+}\mu^{-}\) channel. The bottom subplot in each case shows the relative difference 
(in \%) between the published median expected limit and our recast ULs.
}
    \label{fig:recast_magic_W}
%\end{adjustwidth}    
   \end{figure*}

The key step is to infer the coefficients \(V_i\) from a \emph{pair} of benchmark channels whose ULs are publicly available.  
As a concrete example, we use the MAGIC limits for the dwarf-spheroidal galaxy Coma Berenices~\citep{MAGIC:2021mog}.  
Starting from the published \(b\bar{b}\) limits, we recast them into the \(W^{+}W^{-}\) channel, estimating \(V_i\) once with the \(\tau^{+}\tau^{-}\) channel and once with the \(\mu^{+}\mu^{-}\) channel as the second benchmark (left and right panels of \autoref{fig:recast_magic_W}, respectively).

   \begin{figure*}[h!t]
%\begin{adjustwidth}{-\extralength}{0cm}
    \centering
    \includegraphics[width=0.98\linewidth]{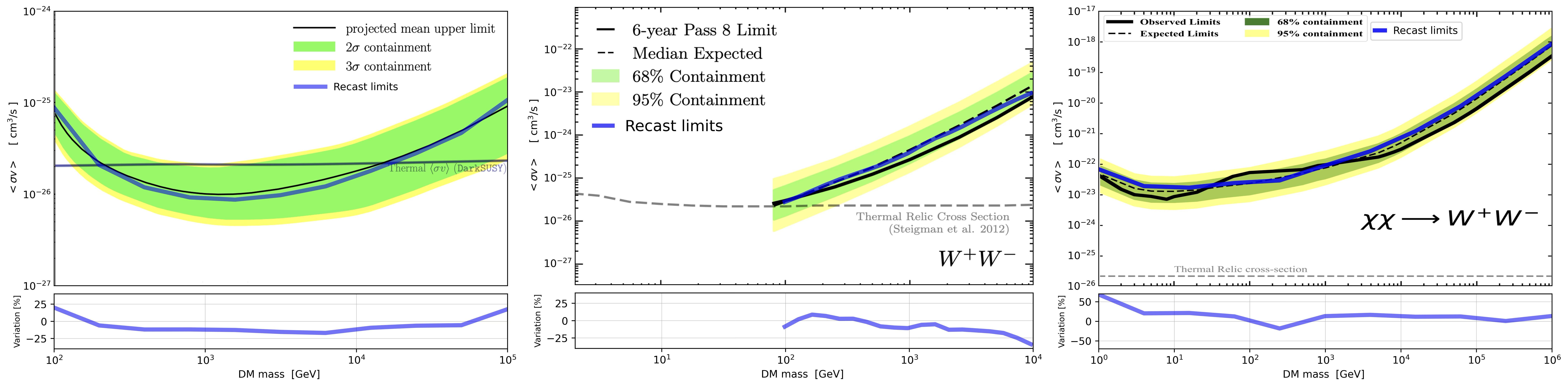}
    \caption{
In both panels, the  image represents the originally published ULs from the respective collaborations. The black line denotes the UL (solid for the observed one, dashed for the median), and the green/yellow shaded regions represent the $68\%$/$95\%$ containment bands of the null hypothesis, as derived from Monte Carlo sampling. 
Overlaid is our recast UL (thick solid blue line) obtained using \autoref{Eq:recast_approx}. 
\textit{Left:} Recast limits for \(W^{+}W^{-}\) annihilation derived from the \(b \bar{b}\) channel using CTAO sensitivity projections for the Galactic Center~\citep{CTA:2020qlo}.  \textit{Center:} Same but with \textit{Fermi}-LAT data on dwarf spheroidal galaxies~\citep{Fermi-LAT:2015att}. In both cases coefficients \(V_i\) inferred using the \(\tau^{+}\tau^{-}\) channel as a second benchmark. 
\textit{Right:} Same but with LHAASO data on dwarf spheroidal galaxies~\citep{LHAASO:2024upb}.  The bottom subplot in each case shows the relative difference 
(in \%) between the published median expected limit and our recast ULs.
}
    \label{fig:recast_exp}
%\end{adjustwidth}
   \end{figure*}

\autoref{fig:recast_magic_W} overlays the recast ULs (thick blue curves) on the original MAGIC figure (Figure 3 of Ref. \citep{MAGIC:2021mog}) .  
The median and uncertainty band ~\citep{MAGIC:2021mog} (derived from the MC sampling of the null hypothesis) are also displayed, providing a direct comparison. 
In both cases, the recast curves track the published median (with a variation smaller than $\sim 25\%$)  to within one standard deviation over the full mass range, demonstrating the precision of the method.  
A parallel exercise for recasting into the \(\mu^{+}\mu^{-}\) channel is presented in~\autoref{App:mumu_MAGIC}, where a similar good agreement is found.  
These tests confirm that the \(V_i\)-based approximation provides a robust tool for translating upper limits between annihilation channels when the full IRF is unavailable.

\medskip
To further validate our approach, we now extend the recasting exercise to different gamma-ray instruments.  
\autoref{fig:recast_exp} displays the results of recasting upper limits on the DM annihilation cross section using published data from CTAO~\citep{CTA:2020qlo}, \textit{Fermi}-LAT~\citep{Fermi-LAT:2015att}, and  LHAASO~\citep{LHAASO:2024upb}.  
In the left panel, we recast the CTAO sensitivity to \(b\bar{b}\) annihilation in the Galactic Center~\citep{CTA:2020qlo} into a limit on the \(W^+W^-\) channel.  
In the center panel, we recast \textit{Fermi}-LAT limits from the \(b\bar{b}\) annihilation in dwarf spheroidal galaxies~\citep{Fermi-LAT:2015att} into the \(W^+W^-\) channel. Lastly, in the right panel we did the same, but for limits obtained by LHAASO ~\citep{LHAASO:2024upb} over 700 days of observations on dwarf spheroidal galaxies.
In all cases, the recast ULs for the \(W^+W^-\) channel are overlayed with originally published limits (with a variation smaller than $\sim 25\%$ for CTA and \textit{Fermi}-LAT, $\sim 50\%$ for LHAASO)  and lie well within the reported \(1\sigma\) containment bands.  
These results reinforce the generality and reliability of our IRF-independent recasting method across a variety of instruments and observational targets.

%while the right panel illustrates the recast limits for \(\tau^+\tau^-\to W^+W^-\) using LHAASO data over 700 days~\citep{LHAASO:2024upb}. 

%\subsection{Casting Specific Microscopic DM Models}

\subsection{Recasting Annihilation Limits into Decay Constraints}
\label{subsec:recast_decay}

The differential gamma-ray flux produced by the \textit{decay} of DM particles $\chi$ is given by
\begin{equation}\label{eq:dm_flux_th_dec}
\frac{d\Phi}{dE}(E) = J_{\rm dec} \cdot \left(\frac{1}{4\pi\, m_{\chi}\,\tau} \frac{dN_\gamma}{dE}\right),
\end{equation}
where the astrophysical $J$-factor for decay is
\begin{equation}\label{eq:j_dec}
J_{\rm dec} \equiv \int_{\Delta\Omega} d\Omega \int_{l.o.s.} dl\,\rho_{\chi}(l, \theta),    
\end{equation}
and $\tau$ denotes the DM decay lifetime. The spectral shape $dN_\gamma/dE$ is analogous to the case of annihilation, except that the energy cutoff now occurs at $E = m_{\chi}/2$.

Consequently, the formalism for decaying DM is nearly identical to that for annihilating DM, with two main substitutions: $m_{\chi} \rightarrow m_{\chi}/2$ in the kinematics, and the appropriate change in the $J$-factor. The expected signal counts per bin become
\begin{align}
    \label{eq:k_i_decay}
    s_i = y \cdot K_i, \quad \text{where}
\end{align}
\begin{align}
    K_i \equiv \int_{\Delta E'_i} dE' \int dE\, A_{\gamma,\mathrm{eff}}(E)\, \mathcal{G}(E, E')\, \frac{dN_\gamma}{dE}\, \frac{T_{\mathrm{obs}}\, J_{\rm ann}}{8\pi\, m_\chi^2},
\end{align}
and
\begin{equation}
    y \equiv \frac{m_{\chi}}{\tau} \cdot \frac{J_{\rm dec}}{J_{\rm ann}},
    \label{eq:y_tau}
\end{equation}
with $J_{\rm ann}$ the canonical annihilation $J$-factor.

Comparing Eqs.~\ref{eq:k_i_decay} and~\ref{eq:k_i}, we see that setting an UL on $y$ is mathematically equivalent to setting a UL on the annihilation cross section $\sigma$. Therefore, from Eq.~\ref{eq:y_tau}, the UL on the DM lifetime is simply
\begin{equation}
    \tau^{\rm UL} = \frac{J_{\rm dec}}{J_{\rm ann}} \cdot \frac{m_\chi}{\sigma^{\rm UL}},
    \label{eq:tau_sigma}
\end{equation}
where $\sigma^{\rm UL}$ is the corresponding upper limit on the annihilation cross section obtained using the annihilation formalism.

\autoref{eq:tau_sigma} holds when spatial morphology is treated identically (spectral-only case) and  as long as the spatial morphology of the DM source is assumed to be the same in both cases, and the additional contribution to the $J-$factor from halo substructures is also neglected~\citep[see, e.g.,][for an estimation of contribution of halo substructures in different classes of targets]{Moline2017SubhaloProps}. However, in general, for extended sources, the projected spatial profile differs between annihilation ($\propto \rho^2$) and decay ($\propto \rho$), and this difference should be taken into account in a complete analysis.

The equivalence in  \autoref{eq:tau_sigma} allows the direct application of all forecasting and recasting techniques developed in this work to decaying DM scenarios, by translating the final constraint on $\sigma$ into one on $\tau$ via Eq.~\ref{eq:tau_sigma}. This holds 

To validate ~\autoref{eq:tau_sigma}, we use the same published CTAO ULs for the dwarf spheroidal galaxy Draco~I as discussed in \autoref{sec:results_forecast} on the validation of the forecasting method. This time, instead of forecasting the annihilation limits, we recast them into decay lifetime limits using ~\autoref{eq:tau_sigma}.

\begin{figure}[htb]
    \centering
    \includegraphics[width=0.85\linewidth]{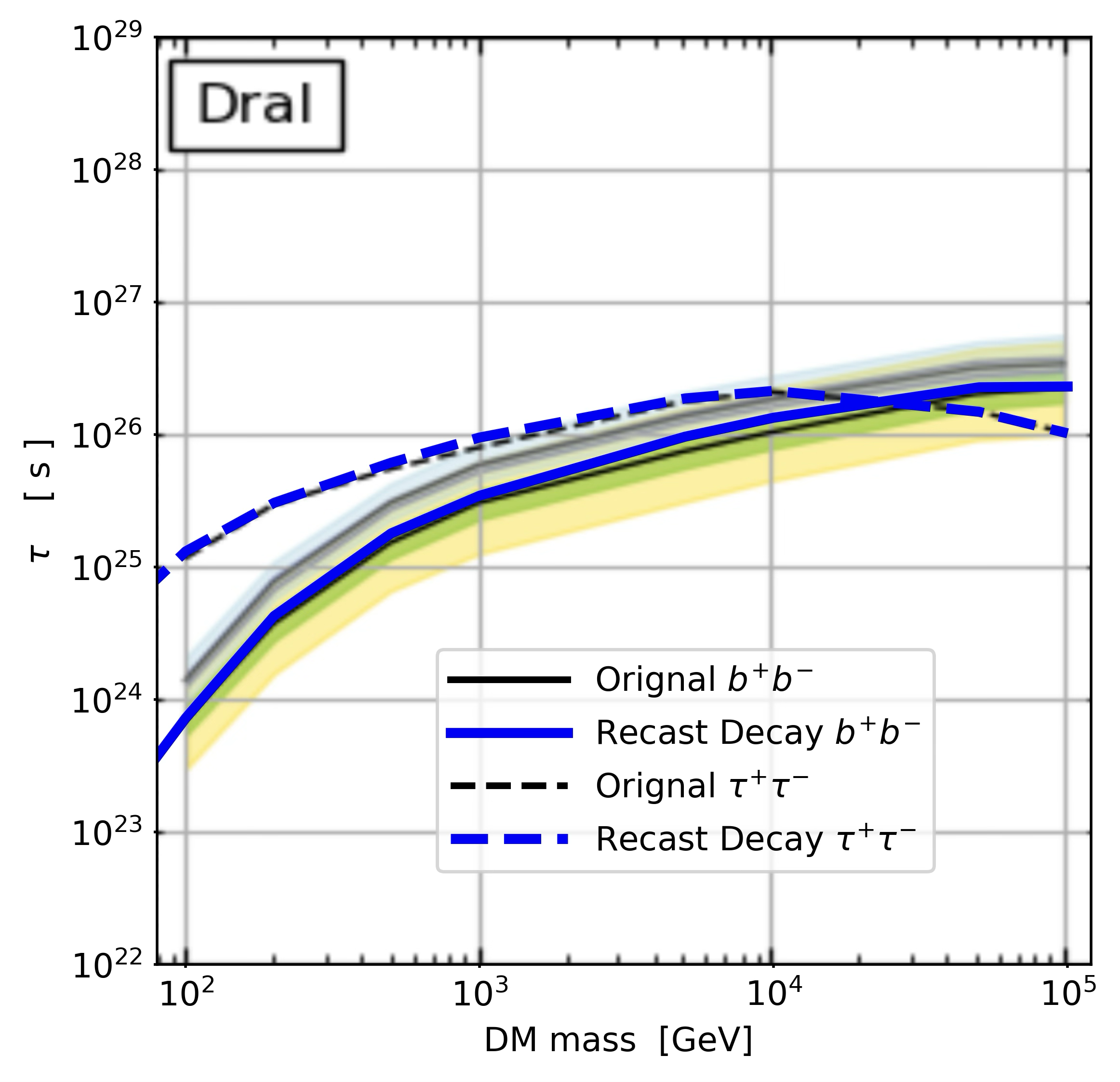}
    \caption{Comparison of recast DM decay lifetime upper limits (solid and dashed blue lines for $b\bar{b}$ and $\tau^+\tau^-$ channels) using Eq.~\ref{eq:tau_sigma}, obtained from the CTAO Draco I annihilation ULs, with the published decay limits from Ref.~\cite{CTAO-dphs}.}
    \label{fig:draco_decay}
\end{figure}

The results are shown in~\autoref{fig:draco_decay}, where our recast decay limits (solid and dashed blue lines for the $b\bar{b}$ and $\tau^+\tau^-$ channels, respectively) are overlaid on the published CTAO decay limits, figure 9 from Ref.~\cite{CTAO-dphs}. The near perfect agreement across the entire range of DM mass provides a direct validation of Eq.~\ref{eq:tau_sigma} and confirms the robustness of our formalism when also applied to the DM decay scenario.

\subsection{Recasting Upper Limits for Higgsino-like and Loop-corrected Annihilation Spectra}
\label{sec:results_cast_higgsino}

Having validated the robustness of our recasting methodology across different experiments and annihilation channels, we now apply this approach to reinterpret already published CTAO ULs, originally computed for the \(W^+W^-\) annihilation channel, for two different cases: 

\bigskip\noindent
\textbf{Higgsino-like scenario:} 
    As anticipated in \autoref{sec:dm_signal_model}, the thermal Higgsino DM, with a mass close to 1.1~TeV, is a well-supported candidate in the search for DM. The Higgsino represents a simple form of weakly interacting massive particle (WIMP) DM, interacting with the Standard Model (SM) through electroweak \(SU(2)_L \times U(1)_Y\) gauge interactions. It belongs to an electroweak multiplet that contains a neutral component, which becomes the DM particle after electroweak symmetry breaking. As a \(SU(2)_L\) doublet, thermal Higgsino DM is too heavy to be produced at current colliders and interacts too weakly to be observed in direct detection experiments. Its annihilation channels include tree-level processes into \(W^+W^-\) and \(Z^+Z^-\) pairs, yielding a broad gamma-ray spectrum extending from the DM mass down to lower energies, as well as one-loop processes that produce a narrow gamma-ray line from the \(\gamma\gamma\) final state, which was shown in \autoref{fig:dm_spectra}. Additionally, one-loop processes involving higher-body final states, such as \(WW\gamma\), generate a sharp endpoint feature slightly below the DM mass, which is nearly indistinguishable from the primary gamma ray line given the energy resolution of current and proposed detectors.
    All this leads to a spectrum similar -- but not identical -- to the pure \(W^+W^-\) case.

    It is important to note that the Higgsino spectrum consists of both a broad 
continuum, dominated by $W^+W^-$ and $ZZ$ final states, and sharp line-like 
features from loop-induced processes such as $\gamma\gamma$, $Z\gamma$, and 
endpoint contributions (see again \autoref{fig:dm_spectra}). 
These line-like terms become especially relevant at multi-TeV masses. 
In the present work our goal is to illustrate the applicability of the recast 
procedure to a non-standard DM scenario, rather than to provide a complete 
phenomenological study of the Higgsino. For this reason we focus on comparing 
with published $W^+W^-$ limits, which serve as a representative continuum 
contribution, while noting that a dedicated analysis including the line-like 
channels would be more appropriate for TeV-scale Higgsinos.

\begin{figure*}[h!t]
 %\begin{adjustwidth}{-\extralength}{0cm}
    \centering
    \includegraphics[width=0.8\linewidth]{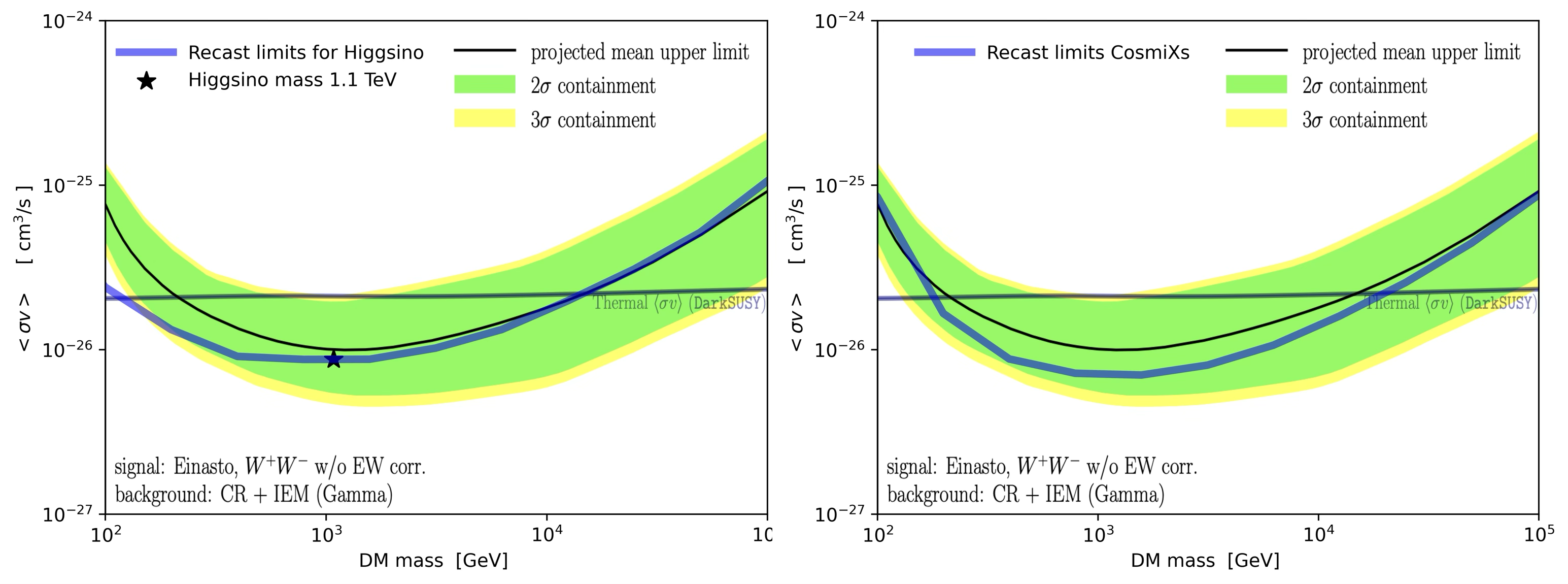}
 \caption{
Recasting CTAO upper limits (ULs) on the DM annihilation cross section into constraints on specific microscopic DM models using \autoref{eq:ratio}. \textit{Left Panel:} Recast ULs (blue line) for the thermal Higgsino model are derived from the CTAO \(b\bar{b}\) ULs~\citep{CTA:2020qlo}. The canonical Higgsino mass of 1.1~TeV is highlighted with a black star, while the recasting extends the constraints over a range of masses. \textit{Right Panel:} Same but assuming the \texttt{cosmiXs}-based DM spectra for the $W^+W^-$ annihilation channel. In both plots the background figure is the same as in the left panel of \autoref{fig:recast_exp}, i.e. the 
CTAO sensitivity projections for the Galactic Center ~\citep{CTA:2020qlo} for the 
$W^+W^-$ annihilation channel.}
    \label{fig:recast_higgsino_cosmixs}
%\end{adjustwidth}    
   \end{figure*}

     We apply \autoref{Eq:recast_approx} to recast original CTAO \(W^+W^-\) limits for the Higgsino DM. The procedure is the same used for producing the left plot of \autoref{fig:recast_exp}. For this spectral type, we compute the corresponding intrinsic photon yield \(\Delta N_{\gamma, i}\) in each energy bin and perform the recasting assuming the same \(V_i\) coefficients generated in \autoref{sec:results_cast_benchmarks}. The resulting ULs are shown in \autoref{fig:recast_higgsino_cosmixs} overlayed to the ``original'' \(W^+W^-\) limits of \citep{CTA:2020qlo} ULs in \autoref{fig:recast_higgsino_cosmixs}. Instead of just considering the preferred Higgsino mass $m_h=1.1$~TeV, marked with a star in the figure, we extend the analysis to all masses, keeping the same Higgsino annihilation branching ratios as in \autoref{eq:dnde}.  One can see that the Higgsino limits overlap those of the \(W^+W^-\) channel for large DM masses, whereas they are more stringent for low DM masses. This is explained in terms of the additional contribution of the line-like photon yield (see again \autoref{fig:dm_spectra}), which if integrated over the entire energy range, contribute fractionally more for lower mass DM\footnote{We note that in the presence of line-like signal a specific search for such smoking gun signatures, e.g. with smaller energy window, would be more appropriate, as done by \citet{Rodd:2024qsi}. However, this is not the aim of this demonstration. Furthermore, we do not include the Sommerfeld enhancement \cite[see, e.g.,][]{Blum:2016nrz}, so our results cannot be directly compared to those of Higgsino detectability as in \citet{Rodd:2024qsi}. A similar argument could be made for the claims of Wino detection in \cite{Safdi:2025sfs}.}.

\bigskip\noindent
\textbf{Recast for \texttt{cosmiXs}-based DM photon yields} 
In \citet{arina2024cosmixs}, the authors revisited the computation of source spectra from dark matter (DM) annihilation and decay by employing the Vincia shower algorithm within Pythia, incorporating QED and QCD final-state radiation as well as electroweak (EW) corrections with massive bosons, which are absent in the default Pythia shower. They included spin correlations and off-shell effects throughout the EW shower and tuned both Vincia and Pythia parameters to LEP data on pion, photon, and hyperon production at the Z pole. Their results were tabulated similarly to those produced in PPPC~\citep{Cirelli:2005uq}, which was previously used in this work. In their work, the authors already compared the \texttt{cosmiXs}-based spectra with the PPPC-based spectra. We reported an example of such comparison in \autoref{fig:dm_spectra} for the $W^+W^-$ channel. It is therefore interesting to see how the experimental limits are affected by this novel, more accurate estimation of the DM photon yield. To do this, we again make use of \autoref{Eq:recast_approx}, this time leaving intact the effective area and modifying the DM spectrum. The results are shown in \autoref{fig:recast_higgsino_cosmixs} (right). One can see that \texttt{cosmiXs}-based limits are stronger than the limits based on PPPC mainly in the central DM mass range \citep[see, e.g.][Fig.~14]{arina2024cosmixs}. This is due to the relative contribution of loop corrections to the total DM flux, which are mass-dependent. We therefore again demonstrate the utility of our algorithm.

\section{Comments and Conclusions}
\label{sec:conclusion}

%This is true in case of velocity-independent DM interactions, and should therefore be modified if e.g. the Sommerfeld effect is included \citep{ark09,Feng:2010zp}.

Throughout this study, we have demonstrated the utility and effectiveness of recasting ULs of DM annihilation cross sections across various theoretical models using existing experimental results. Applying \autoref{eq:ratio}, we were able to reinterpret ULs for a range of DM annihilation channels and for different instrumental setups. This recasting approach has proven particularly beneficial in expanding the constraints on DM properties beyond the specific scenarios originally analyzed, without the need for new observational data.

Our analyses have underscored the importance of making experimental details such as the effective area \( A_{\gamma,\mathrm{eff}}(E) \) publicly available alongside the published ULs. Access to these details greatly enhances the transparency and reproducibility of the results and facilitates the application of recasting techniques. Therefore, we strongly recommend that future publications of DM ULs include not only the limits themselves but also the associated effective areas used in the UL computations. Providing such information would enable the broader scientific community to efficiently apply these results to alternative models and scenarios, thereby maximizing the scientific return from the experimental data\footnote{We also mention that, alternatively, the \textit{Fermi}-LAT collaboration published some dark matter limits~\citep[see, e.g.,][]{Fermi-LAT:2015att} along with the energy-binned likelihood profile test statistics on the signal flux. Although this approach in principle enables recasting of \textit{Fermi}-LAT results, its applicability remains restricted to this experiment, since such likelihood profiles are not generally released by other instruments. In contrast, our method is more general, as it can be applied to any published upper limits, regardless of whether likelihood profiles are made available. }.

Moreover, our work highlights the robustness of the recasting methodology, confirming that it can provide accurate estimates of DM constraints even when transitioning between different annihilation channels or experimental setups. This robustness is crucial for the ongoing efforts to understand the nature of DM and constrains properties across a diverse landscape of theoretical models.

In conclusion, the ability to recast ULs using the methodology outlined in this work offers a powerful tool for the DM research community. It allows for a broader exploration of DM parameter space and helps in pinpointing the characteristics of DM interactions with the Standard Model. As future experiments continue to improve in sensitivity and as new data become available, the recasting approach will remain an essential component in the quest to uncover the nature of dark matter.

\paragraph{Author Contribution}
MD conceived the project. GDA and MD developed the methodology. GDA was responsible for software implementation. Validation was carried out by GDA, MD and MDC (see also \citep{DeCaria_2023}). The original draft was prepared by GDA and MD, with all authors contributing to review and editing. All authors have read and approved the final version of the manuscript.

\paragraph{Data Availability}
Supplementary data and code to reproduce the results of this work can be found in~\citep{zenodo}.

\paragraph{Funding} The work of GDA on this project was supported by the Beatriu de Pin\'{o}s program (BP 2023).

\paragraph{Acknowledgments}  
We thank Nicolao Fornengo and  Javier Rico for carefully reading the manuscript and providing valuable comments and suggestions that helped improve this work.  
This research has made use of the CTA instrument response functions provided by the CTA Consortium and Observatory, see \url{https://www.ctao-observatory.org/science/cta-performance/} (version prod5 v0.1; \citep{ctao_irfs}) for more details. This work made use of Gammapy \citep{gammapy:2023}, a community-developed Python package. The Gammapy team acknowledges all past and current contributors, as well as all contributors of the main Gammapy dependency libraries: \hyperref[https://numpy.org/]{NumPy}, \hyperref[https://scipy.org/]{SciPy}, \hyperref[http://www.astropy.org]{Astropy}, \hyperref[https://astropy-regions.readthedocs.io/]{Astropy Regions}, \hyperref[https://scikit-hep.org/iminuit/]{iminuit}, \hyperref[https://matplotlib.org/]{Matplotlib}.

\bibliographystyle{elsarticle-harv} 

\bibliography{biblio}

\onecolumn
\appendix  

%\clearpage
%%%%%%%%%%%%%%%%%%%%%%%%%%%%%%%%%%%%%%%%%%
%\begin{adjustwidth}{-\extralength}{0cm}

%  start

%\section[\  name~\thesection]{}
%\subsection[\  name~\thesubsection]{}

\section{Second Derivative of the Log-Likelihood}
\label{App:Second_Der}

We derive here the second derivative of the function defined in Eqs.~\autoref{eq:cash_f} and~\autoref{Eq:Lklratio_b2}, which we recall below:

\begin{align}
    f(s) &= s - n \ln(s + b) + C, \quad \text{(C-statistic)} \\
    f(s) &= s - n \ln(s + b) + (1 + \alpha) b - m \ln(\alpha b) + C, \quad \text{(On/Off statistic)}
\end{align}

\subsection*{C-statistic case}

In the first case, the background \( b \) is known and is therefore treated as a fixed parameter. The first and second derivatives of the log-likelihood function are then:

\begin{align}
    f'(s) &= 1 - \frac{n}{s + b} \sim 0, \\
    f''(s) &= \frac{n}{(s + b)^2} \sim \frac{1}{b}
\end{align}

where the approximation on the right-hand side holds under the null hypothesis \( s = 0 \), for which the expected value of \( n \) is \( b \). Throughout this section, the symbol “\(\sim\)” denotes evaluation under the null hypothesis.

\subsection*{On/Off statistic case}

In the On/Off case, \( b \) is a nuisance parameter and depends on \( s \) by profiling. That is, for each value of the signal strength \( s \), the background \( b \) is taken to be the one that maximizes the likelihood. It can be shown~\cite{Cowan2011,DAmico:2022psx} that this maximum occurs at:

\begin{equation}
    b(s) = \frac{n_1(s) + n_2(s)}{2 (1 + \alpha)},
\end{equation}

where we define:

\begin{align}
    n_1(s) &= n + m - (1 + \alpha) s, \\
    n_2(s) &= \sqrt{n_1^2(s) + 4 (1 + \alpha) s m}.
\end{align}

The first and second derivatives of \( b(s) \) with respect to \( s \) are:

\begin{align}
    \frac{db}{ds} &= \frac{2m - n_1 - n_2}{2 n_2} \sim -\frac{1}{1 + \alpha}, \\
    \frac{d^2b}{ds^2} &= \frac{(1 + \alpha^{-1})(n_1 + n_2 - 2m)(n_2 + 2m - n_1)}{2 \alpha^{-1} n_2^3} \sim \frac{2 \; \alpha}{(1 + \alpha)^2 b},
\end{align}

where in both expressions, the approximations again correspond to taking the expectation under the null hypothesis, \( s = 0 \), so that \( n = b \) and \( m = \alpha b \).

At this point, one can write down the first and second derivatives of $f$ as follows:

\begin{align}
    f'(s) &=  - \frac{n }{s + b} \left(1 + \frac{db}{ds}\right)
- \frac{m}{b} \frac{db}{ds} + 1 + (1 + \alpha) \frac{db}{ds} \sim 0, \\
    f''(s) &= n \frac{(1 + \frac{db}{ds})^2 - (s + b) \frac{d^2 b}{ds^2}}{(s + b)^2}
+ m \frac{\left( \frac{db}{ds} \right)^2 - \frac{d^2 b}{ds^2} b }{b^2}
+ (1 + \alpha) \frac{d^2 b}{ds^2}  \sim \frac{1}{b(1+\alpha^{-1})}.
\end{align}

\section{Derivation of the Inequality \autoref{eq:inequality}}\label{App:Inequality}

To demonstrate the inequality in \autoref{eq:inequality}, we invoke the Cauchy--Schwarz inequality. Define two vectors with positive components:
\begin{equation}
    X_i = \frac{K_i}{\sqrt{b_i}}, \quad Y_i = \sqrt{b_i}.
\end{equation}

Applying the Cauchy--Schwarz inequality, we have the following.
\begin{equation}
    \left(\sum_i X_i Y_i\right)^2 \leq \left(\sum_i X_i^2\right)\left(\sum_i Y_i^2\right).
\end{equation}

Substituting $X_i$ and $Y_i$ back, we obtain:
\begin{equation}
    \left(\sum_i K_i\right)^2 \leq \left(\sum_i \frac{K_i^2}{b_i}\right)\left(\sum_i b_i\right).
\end{equation}

Rearranging terms gives precisely the stated inequality:
\begin{equation}
    \sqrt{\sum_i \frac{K_i^2}{b_i}} \geq \frac{\sum_i K_i}{\sqrt{\sum_i b_i}}.
\end{equation}

\section{The factors $V_i$ are approximately model independent}\label{App:Vi}

In this section, we show that the coefficients \( V_i \) introduced in \autoref{Eq:Vi_main_text},

\begin{equation}
    V_i \equiv \frac{K_i}{\sqrt{b_i} \, \Delta N_{\gamma,i}} 
    = \frac{1}{\sqrt{b_i} \, \Delta N_{\gamma,i}} \int_{\Delta E'_i} dE' \int dE \, A_{\gamma,\mathrm{eff}}(E) \cdot \mathcal{G}(E, E') \cdot \frac{dN_{\gamma}}{dE_{\gamma}},
    \label{Eq:V_i}
\end{equation}

are approximately model-independent under reasonable assumptions. In particular, they depend primarily on the instrument response functions (IRFs), rather than on the specific shape of the DM gamma-ray spectrum.

Since the integrals in \autoref{Eq:V_i} are computed numerically over discrete bins, we approximate them as:

\begin{equation}
    V_i \simeq \frac{\Delta E'_i}{\sqrt{b_i} \, \Delta N_{\gamma,i}} \sum_j A_j \cdot \mathcal{G}_{ij} \cdot \frac{ \Delta N_j }{ \Delta E_j },
\end{equation}

where \( A_j \) is the effective area averaged over the bin \( j \), \( \mathcal{G}_{ij} \) is the energy dispersion matrix element from the true bin \( j \) to the reconstructed bin \( i \), and \( \Delta N_j \) is the intrinsic photon count in the bin \( j \).

Rewriting this expression yields the following:

\begin{equation}
    V_i \simeq \frac{1}{\sqrt{b_i}} \left( A_i \cdot \mathcal{G}_{ii} + \sum_{j \neq i} A_j \cdot \mathcal{G}_{ij} \cdot \frac{ \Delta N_j / \Delta E_j }{ \Delta N_i / \Delta E'_i } \right).
    \label{Eq:V_i_bins}
\end{equation}

From this we see that \( V_i \) is approximately given by:

\begin{equation}
    V_i \approx \frac{1}{\sqrt{b_i}} \, A_i \cdot \mathcal{G}_{ii},
\end{equation}

provided that the following condition holds:

\begin{equation}
    \frac{ \mathcal{G}_{ij} \cdot \Delta N_j / \Delta E_j }{ \Delta N_i / \Delta E'_i } \ll 1 \quad \text{for all } j \neq i.
\end{equation}

This condition is typically satisfied when the instrument has good energy resolution and the spectrum \( dN_{\gamma}/dE \) varies slowly compared to the bin width.

\begin{figure*}[ht]
%\begin{adjustwidth}{-\extralength}{0cm}
    \centering
  \includegraphics[width=0.8\linewidth]{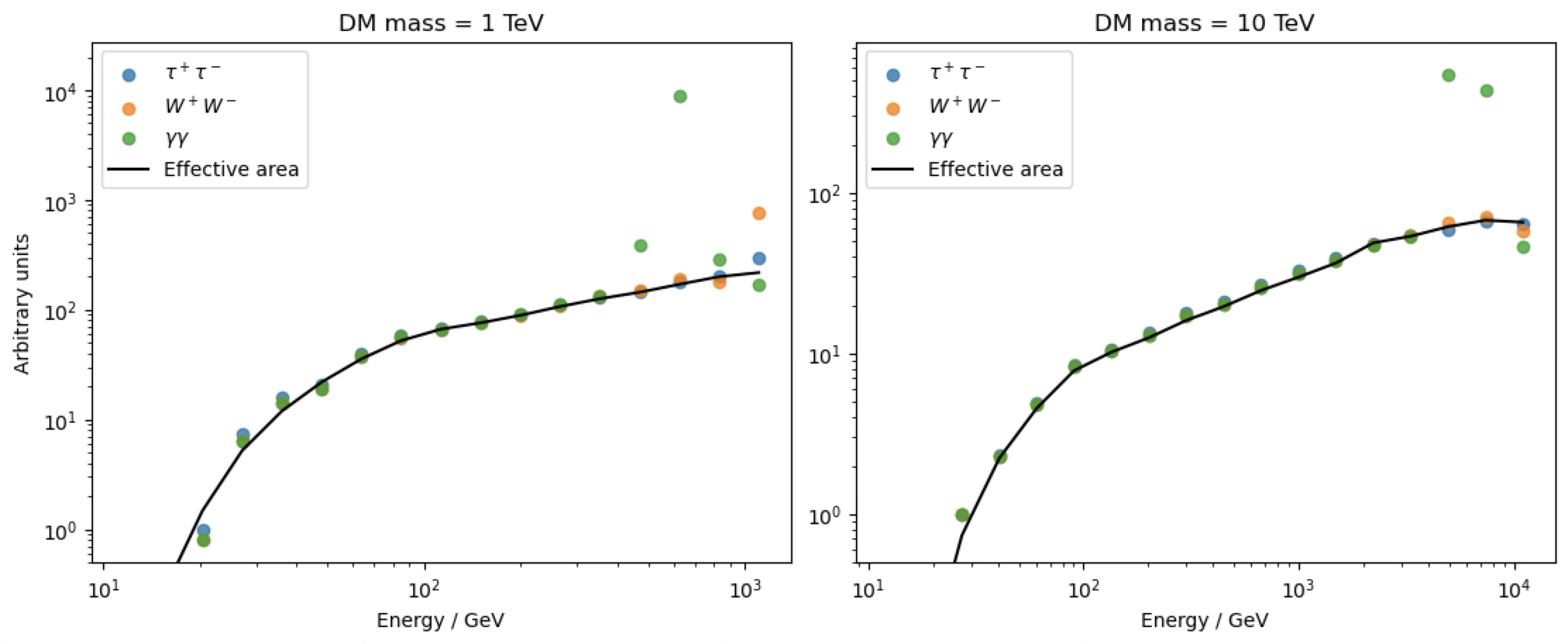}
    \caption{
Comparison of the scaling factors \( K / \Delta N_{\gamma} \) across different dark matter (DM) annihilation channels—\(\tau^+\tau^-\), \(W^+W^-\), and \(\gamma\gamma\)—for two benchmark DM masses: 1~TeV (left) and 10~TeV (right). Each colored point represents the value of \( K / \Delta N_{\gamma} \)  in a given reconstructed energy bin, normalized to arbitrary units. The black line shows the shape of the effective area \( A_{\gamma,\mathrm{eff}} \) for reference. For smoothly varying spectra (\(\tau^+\tau^-\), \(W^+W^-\)), the \(  K / \Delta N_{\gamma} \)  values closely follow the effective area profile, confirming their approximate model independence. In contrast, the \(\gamma\gamma\) channel—featuring a sharp spectral line—exhibits significant deviations at energies close to the DM mass, illustrating the breakdown of the model-independence approximation in this regime.
}
    \label{fig:Vi}
%\end{adjustwidth}    
   \end{figure*}

We have numerically tested this approximation using various DM annihilation channels and the CTAO IRFs~\cite{ctao_irfs}. The resulting values of \( K / \Delta N_{\gamma} \) agree within approximately 10\% across different DM models\footnote{An exception is the case of DM annihilation to $\gamma\gamma$, 
where the spectrum $dN_{\gamma}/dE$ has a very sharp peak. In such cases, 
the rapid variation within a single energy bin violates the assumptions 
underlying this approximation, rendering it inaccurate. 
It is also worth noticing that electroweak (EW) corrections generate 
bump-like features near $E \simeq m_\chi$ (notably for the $W^+W^-$ channel), 
which can similarly reduce the accuracy of a single $V_i$ set. 
Our numerical tests (see \autoref{fig:Vi}) show that this degradation is localized 
near these endpoints.}. The results of this comparison are shown in \autoref{fig:Vi}, where \( K / \Delta N_{\gamma} \)  values are plotted across energy bins and DM masses for different annihilation channels.

\section{Recasting MAGIC UL for the $\mu^+\mu^-$  Channel} \label{App:mumu_MAGIC}
\begin{figure*}[h!t]
%\begin{adjustwidth}{-\extralength}{0cm}
    \centering
    \includegraphics[width=0.8\linewidth]{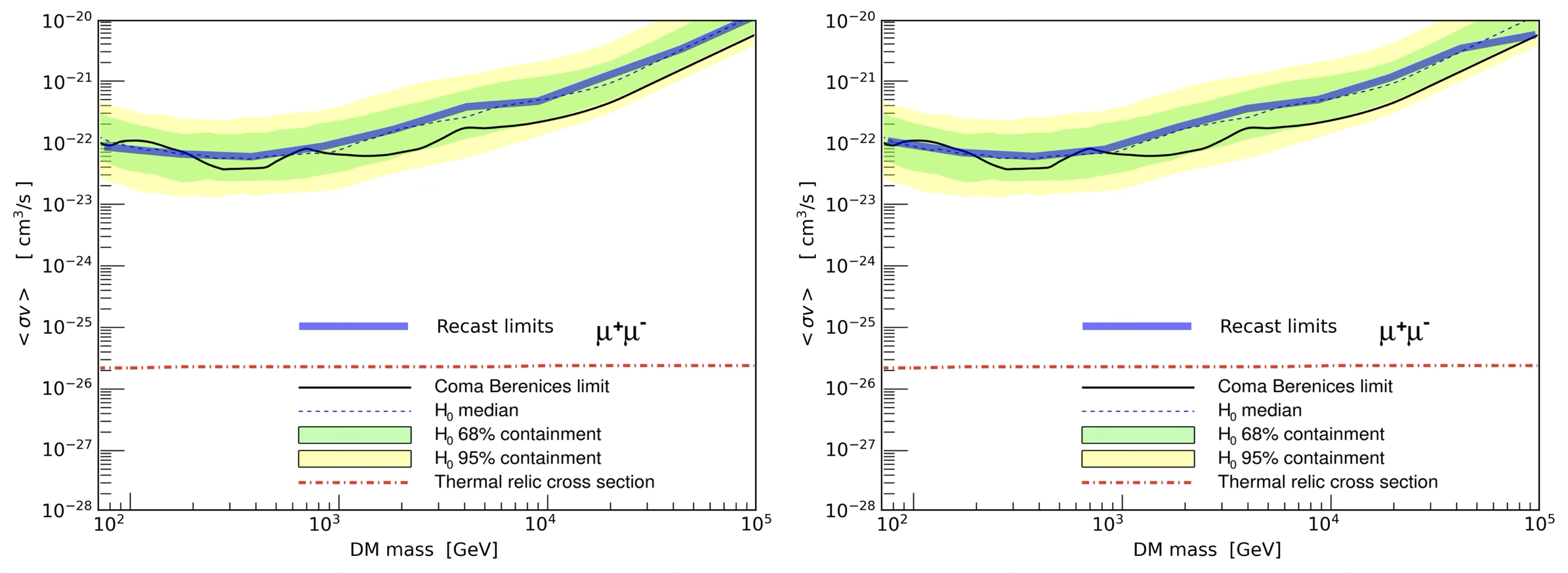}
    \caption{
In both panels,  the black solid line represents the originally published  ULs from the MAGIC collaboration~\citep{MAGIC:2021mog} for the \(\mu^{+}\mu^{-}\) annihilation channel. The black dotted line denotes the median and the green/yellow shaded regions denote the $68\%$/$95\%$ band of the null hypothesis,  as derived from Monte Carlo sampling.
Overlaid is our recast UL (thick solid blue line) obtained using \autoref{Eq:recast_approx} to recast limits for \(\mu^{+}\mu^{-}\) annihilation from the \(\tau^{+}\tau^{-}\) channel. \textit{Left:} ULs obtained using coefficients \(V_i\) inferred using the \(b \bar{b}\) channel as a second benchmark.  
\textit{Right:} ULs obtained using coefficients \(V_i\) inferred using the \( W^{+} W^{-}\) channel. }
    \label{fig:recast_magic_mu}
%\end{adjustwidth}    
   \end{figure*}

To illustrate that the IRF-free recasting procedure works equally well for other final states, we repeat the exercise of~\autoref{sec:results_cast_benchmarks} for the \(\mu^{+}\mu^{-}\) annihilation channel.  
The steps are identical:

\begin{enumerate}
    \item We adopt the MAGIC ULs on the \(\tau^{+}\tau^{-}\) channel in Coma Berenices as the primary input.  
    \item The coefficients \(V_i\) in \autoref{Eq:recast_approx} are inferred from a second benchmark channel whose ULs are also published by MAGIC.  
          Two choices are considered:
          \begin{itemize}
              \item the \(b\bar{b}\) channel (left panel of \autoref{fig:recast_magic_mu});
              \item the \(W^{+}W^{-}\) channel (right panel of \autoref{fig:recast_magic_mu}).
          \end{itemize}
    \item Using the extracted \(V_i\) array, we recast the limits \(\tau^{+}\tau^{-}\) into the \(\mu^{+}\mu^{-}\) channel through \autoref{Eq:recast_approx}.
\end{enumerate}

\autoref{fig:recast_magic_mu} compares the recast ULs (thick blue curves) with the published MAGIC ULs (black solid curves).  
As in the main text, the black dotted line and the green / yellow bands indicate the median and \(68\%/95\%\) containment of the null-hypothesis pseudo-experiments reported by MAGIC.  
In both benchmark configurations, the recast limits track the published median to within one standard deviation over the full mass range, reinforcing the robustness of the \(V_i\)-based approximation.

\end{document}